\documentclass[amsmath,amssymb,11pt]{article}
\pdfoutput=1 

\usepackage{jheppub2} 
\usepackage{tikz}

\usepackage[T1]{fontenc} 
\definecolor{mydarkgreen}{RGB}{0,100,0}
\usepackage[numbers]{natbib}
\usepackage[T1]{fontenc} 
\usetikzlibrary{decorations.pathmorphing}
\usepackage{pgfplots}
\pgfplotsset{compat=1.17}
\usepackage{comment}
\usepackage{subcaption}
\usepackage{extarrows}
\usepackage{epsfig}
\usepackage{braket}
\usepackage{ulem}
\usepackage{graphicx}
\usepackage{float}
\usepackage{dsfont}
\usepackage{subcaption}
\usepackage{amsmath}
\usepackage{amsmath,amsopn}
\usepackage{xcolor }
\usepackage{bbold}
\usepackage{tikz}
\usepackage{mathrsfs}
\usepackage{amsmath}
\usepackage{amssymb}
\usepackage{color}
\usepackage[utf8]{inputenc}
\usepackage{graphicx}
\usepackage{bm}
\usepackage{mathrsfs} 
\usepackage{braket}
\usepackage{dsfont}
\usepackage{tikz}
\usepackage{tikz}
\usetikzlibrary{decorations.pathmorphing, arrows.meta}

\usetikzlibrary{arrows}
\usetikzlibrary{decorations.markings, decorations.pathmorphing}

\usetikzlibrary{arrows}

\usepackage{extarrows}
\usepackage{epsfig}
\usepackage{braket}
\usepackage{ulem}
\usepackage{graphicx}
\usepackage{float}
\usepackage{dsfont}
\usepackage{subcaption}
\usepackage{amsmath}
\usepackage{amsmath,amsopn}
\usepackage{xcolor }
\usepackage{bbold}
\usepackage{tikz}
\usepackage{mathrsfs}
\usepackage{amsmath}
\usepackage{amssymb}
\usepackage{color}
\usepackage[utf8]{inputenc}
\usepackage{graphicx}
\usepackage{bm}
\usepackage{mathrsfs} 
\usepackage{braket}
\usepackage{dsfont}
\usepackage{tikz}
\usepackage{tikz}
\usetikzlibrary{arrows}
\usetikzlibrary{decorations.markings, decorations.pathmorphing}

\usetikzlibrary{arrows}

\definecolor{pink1}{rgb}{0.858, 0.188, 0.478}

\newcommand{\beq}{\begin{equation}}

\newcommand{\eeq}{\end{equation}}

\def\hs{\hskip 1pt}

\def\beq{\begin{equation}}
\def\eeq{\end{equation}}
\def\be{\begin{equation}}
\def\ee{\end{equation}}

\title{\boldmath Quasinormal Modes and the Switchback Effect in Schwarzschild-de Sitter
}

\author[a]{Mir Mehedi Faruk,}
\author[b,c]{Facundo Rost,}
\author[b,c]{Jan Pieter van der Schaar.}

\affiliation[a]{Department of Physics, McGill University, Montreal, QC, H3A 2T8, Canada}
\affiliation[b]{Institute of Physics, University of Amsterdam, Science Park 904, PO Box 94485, 1090 GL Amsterdam, the Netherlands}
\affiliation[c]{Delta Institute for Theoretical Physics, Science Park 904, PO Box 94485, 1090 GL Amsterdam, the Netherlands}

\emailAdd{ 
mir.faruk@mail.mcgill.ca, f.e.rost@uva.nl, j.p.vanderschaar@uva.nl }

\abstract{We study the causal structure of Schwarzschild-de Sitter (SdS), including shock wave perturbations, in $D>3$ using reflected null ray trajectories, either through the interior black hole or the exterior de Sitter region. Specifically, we compute the quasinormal mode frequencies in the eikonal, high-frequency, limit, by identifying the `critical time', for arbitrary values of the black hole mass. We emphasize the important role of the static sphere proper time normalization and related boundary conditions. The computed critical times indicate the presence of singularities in the late-time, large mass, scalar field correlator in SdS, which should be resolved by introducing complex geodesics consistent with interior black hole and exterior de Sitter effective thermofield double states. In addition we relate the critical time to a diverging holographic complexity observable and compute the `switchback' delay by adding a pair of shock wave perturbations for arbitrary values of the mass of the black hole.} 

\begin{document} 
\maketitle
\flushbottom

\newpage
\section{Introduction}
Through the AdS/CFT correspondence \cite{Gubser:1998bc,Maldacena:1997re,Witten:1998qj,Almheiri:2019psf,Maldacena:2015waa,Papadodimas:2013jku,Horowitz:1999jd,Festuccia:2005pi,Fidkowski:2003nf} one has made significant progress in understanding many interesting avenues of quantum gravity and black hole physics in recent years. The chaotic features of black hole horizons in AdS spacetime are particularly intriguing, as understood for instance by studying shock waves and out-of-time-ordered correlators (OTOC) \cite{Shenker:2013pqa,Ceplak:2024bja,Shenker:2013yza,Gao:2016bin,Geng:2021wcq,Stanford:2014jda,Dalui:2018qqv,Couch:2016exn,Sekino:2008he,Horowitz:2023ury}. 
Our universe, however, seems to be best described by an early- and late-time phase of accelerated expansion \cite{Peebles:2002gy}, which can be approximated by a de Sitter (dS) instead of an Anti-de Sitter universe. Although the lack of a well-established holographic framework for quantum gravity in dS spacetime restricts the scope of similar studies \cite{Galante:2023uyf,Balasubramanian:2002zh,Bernardo:2021zxo,Kolchmeyer:2024fly,Denef:2009kn}, gravitational bulk notions of complexity and chaos in AdS  \cite{Ryu:2006ef,Carmi:2017jqz,Chapman:2018lsv,Chapman:2018dem,Flory:2018akz,Caceres:2019pgf,Brown:2015lvg,Goto:2018iay,Couch:2016exn} can be generalized and studied in a time-dependent spacetime environment featuring cosmological horizons, which includes de Sitter as the maximally symmetric example \cite{Anegawa:2023dad,Jorstad:2022mls,Baiguera:2023tpt,Chapman:2021eyy,Aguilar-Gutierrez:2024rka,Baiguera:2024xju,Susskind:2021esx,Aguilar-Gutierrez:2023ccv,Ahmed:2020fai,Milekhin:2024vbb,Aguilar-Gutierrez:2023zqm,Aguilar-Gutierrez:2023pnn,Ahmed:2023uem,Ahmed:2020fai}.

An important discrepancy between dS and (black holes in) AdS spacetime is the absence of a spatially asymptotic and non-gravitating boundary theory from which we can probe the static patch \cite{Gao:2000ga,Gao:2016bin,Galante:2023uyf}, preventing the identification of well-defined (gauge-invariant) observables in dS quantum gravity. Proceeding semi-classically, one finds that cosmological horizons exhibit some striking differences as compared to black hole horizons in asymptotic AdS spacetime, for instance when exposed to shock waves \cite{Aalsma:2020aib,Geng:2020kxh,Anninos:2018svg,Aalsma:2021kle,Hayden:2007cs,Hotta:1992wb, Bintanja:2023vel,Hirano:2019ugo}. In particular, geodesics crossing a positive-energy shock wave experience a gravitational time delay in Minkowski and Anti-de Sitter space \cite{Gao:2000ga,Galante:2022nhj}, while they experience a time advance in dS spacetime. A shock wave perturbation satisfying the Null Energy Condition (NEC) in de Sitter space can therefore allow signals from otherwise causally disconnected regions, separated by a cosmological horizon, to be exchanged, opening up a traversable wormhole \cite{Aalsma:2020aib,Geng:2020kxh,Anninos:2018svg}, which for the AdS eternal black hole requires the introduction of NEC violating matter. For the AdS eternal black hole one can effectively describe the exchange of information through a traversable wormhole by introducing boundary operators that give rise to negative energy shock waves in the bulk.
A similar idea can be applied in de Sitter space by introducing positive energy (outgoing) shock waves in a particular static patch region. This is an intrinsic bulk operation that backreacts, it cannot be realized by just imposing different boundary conditions, and as a consequence one is therefore led to study the Schwarzschild-de Sitter (SdS) geometry instead \cite{Aalsma:2020aib}, which due to the appearance of both a cosmological horizon and a black hole horizon is an interesting spacetime in its own right \cite{Draper:2022xzl,Choudhury:2004ph,Qiu:2019qgp,Aalsma:2019rpt,Fernandez-Silvestre:2021ghq,Anderson:2020dim,Faruk:2024usu}.

Recently, we studied the causal structure of SdS black holes with respect to free-falling observers that remain at a fixed and unique distance from the black hole and cosmological horizon, which we referred to as `static-sphere' observers. With respect to the static sphere observers, exterior de Sitter past/future spacelike infinity in the Penrose diagram bends outward. This results in a larger exterior causal diamond as compared to empty dS space. The spacelike singularity of the black hole instead bends inward, resulting in a smaller black hole interior. We computed the outward and inward bending with respect to the static sphere observers for any mass of the black hole, finding that the respective bending first grows to a maximum and then decreases back to the de Sitter perfect square in the Nariai limit \cite{Faruk:2023uzs}. 
 
In this paper we extend the analysis of null geodesics in the SdS geometry and use it to derive the asymptotic (large frequency) quasinormal mode spectrum of a massive scalar field for both the interior black hole and the exterior de Sitter region \cite{Amado:2008hw}. Information about the quasinormal mode spectrum is of particular interest for the SdS partition function and potential holographic descriptions \cite{Denef:2009kn, Susskind:2021esx, Susskind:2021omt, VerlindeH:2023SYKdS, Verlinde:2024znh}. 
In a geodesic WKB approximation of the correlator the quasinormal mode frequencies are related to isolated poles \cite{Konoplya:2011qq,Berti:2009kk,Moitra:2024ixh}, and our results suggest that these are `resolved' in the same way as in the AdS eternal black hole by introducing interior and exterior thermofield double states, leading to a sum over two complex conjugate geodesics in the WKB approximation and moving the singularities to a different Riemann sheet \cite{Fidkowski:2003nf, Chapman:2022mqd, Aalsma:2022eru}. 

Throughout we stress the important role of static sphere observers: the resulting proper time normalization and associated boundary conditions imply that both the flat space and Nariai limits are well-behaved and reproduce standard expectations. It also allows the semi-classical identification of an interior and exterior bulk effective thermofield double state for a massive scalar field, by decoupling the interior black hole and exterior de Sitter region through perfectly reflecting Dirichlet boundary conditions at the static sphere radius, enforcing equilibrium. This also means that the assumed boundary conditions for the quasinormal modes in this paper are different from the boundary conditions assumed in earlier work on SdS \cite{Konoplya:2004uk,Konoplya:2022xid,Konoplya:2024ptj,Sarkar:2023rhp}, which correspond to purely in-going waves at the event horizon and purely out-going waves at the cosmological horizon, instead of Dirichlet boundary conditions at the static sphere. We will assume that imposing Dirichlet boundary conditions at the static sphere, effectively decoupling the interior and exterior part of the SdS geometry, is consistent to leading, semi-classical, order, keeping the spacetime background fixed\footnote{Physically, one could think of a spherical mirror, or brane, with some tension, to ensure that these Dirichlet boundary conditions are stable to leading semi-classical order.}. 

Finally, our results for the SdS causal structure also influence the time-dependent behavior of the volume of the wormhole Wheeler-de Witt (WdW) patch, which is conjectured to be related to holographic quantum complexity \cite{Stanford:2014jda,Couch:2016exn}. By including the effect of a pair of shock wave perturbations, with vanishing total energy, we also compute the switchback time-delay effect for arbitrary mass of the black hole, for both the interior black hole and the exterior de Sitter complexity parameter. Concretely this means we will compute, from the natural perspective of static sphere observers, the so-called `critical times' associated with the geometry in the presence of the pair of shock wave perturbations, with the positive energy shock wave emitted through the cosmological event horizon and the negative energy shock wave sent into the black hole \cite{Anegawa:2023dad, Baiguera:2023tpt, Aguilar-Gutierrez:2023pnn}. We separately study the switch-back delay for the exterior de Sitter, as a consequence of a positive energy shock wave, and a corresponding switch-back advance for the interior black hole region, as a result of a negative energy shock wave. A crucial (and new) aspect of this static sphere critical time calculation, for arbitrary black hole mass in $D>3$, is the influence of the (deformed) causal structure, on top of the standard local shift from the shock waves.

To set the stage and emphasize the general applicability of these techniques, we begin by reminding the reader of critical time calculations for a black hole in AdS spacetime (SAdS\(_D\)), showing how the inward bending of the singularity influences the critical time \cite{Fidkowski:2003nf, Brecher:2004gn, Faruk:2023uzs} in dimensions \(D > 3\). Subsequently, we review and extend this analysis for both the interior black hole and exterior de Sitter region in SdS \cite{Faruk:2023uzs} and apply the results to obtain the quasinormal mode spectrum in the eikonal, high frequency, limit. We then explore the relation between critical time and complexity and introduce shock waves to perturb the SdS geometry, again after first reminding the reader of similar calculations in the context of AdS black holes. Expert readers can skip the AdS technical preliminaries and review sections. Finally, we summarize our results in the conclusions, provide a few final comments and present some remaining open problems. 
 

\section{Technical preliminaries: probing causal structures beyond horizons}

We will only consider spherically symmetric metrics of the following general form in $D=d+1$ dimensions
\begin{equation}
    ds^2=-f(r)dt^2+\frac{1}{f(r)}dr^2+ r^2 \, d\Omega^2_{D-2}\,,\label{general black}
\end{equation}
where $d\Omega_{D-2}$ is the line element of a unit sphere in $D-2$ dimensions. Eventually we will also introduce a radial shock wave on the SdS black hole background, but for now we just want to remind the reader of some basic methods and results that probe the detailed causal structure of Schwarzschild-Anti-de Sitter (SAdS), de Sitter (dS) and Schwarzschild-de Sitter (SdS) geometries. 

\subsection{Black holes in Anti-de Sitter space}	
\label{Kruskal coordinates of AdS blackhole spacetime}

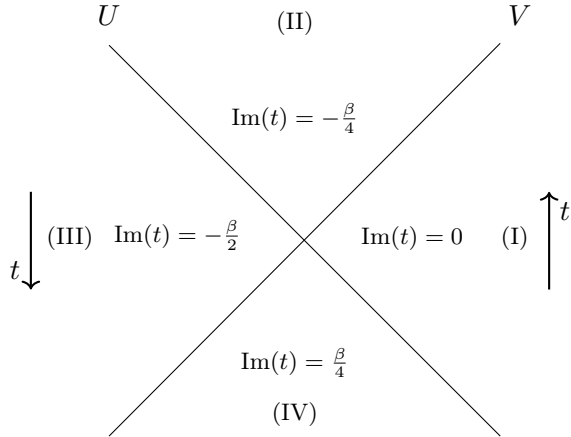
\begin{figure}[t!]
    \centering
    \begin{tikzpicture}[scale=1.3]

        \draw (0,0) -- (4,4.02);
        \draw (0,4) -- (4,0);
           \node[above, font=\footnotesize] at (1.9,4) {(II)};
    \node[above, font=\footnotesize] at (4.15,1.8) {(I)};
    \node[above, font=\footnotesize] at (-0.40,1.8) {(III)}; 
    \node[above, font=\footnotesize] at (1.9,0.0) {(IV)};
        \node[above] at (4.2,4.1) {$V$};
        \node[above] at (0,4.1) {$U$};
        \node[above, font=\footnotesize] at (1.9,3) {Im$(t)=-\frac{\beta}{4}$};
        \node[above, font=\footnotesize] at (3.1,1.8) {Im$(t)=0$};
    \node[above, font=\footnotesize] at (1.9,0.5) {Im$(t)=
    \frac{\beta}{4}$};
      \node[above, font=\footnotesize] at (0.7,1.8) {Im$(t)=
   - \frac{\beta}{2}$};   \draw[->,thick] (4.5,1.5) -- (4.5,2.5) node[pos=0.8,right]{$t$};
     \draw[->,thick]  (-0.8,2.5) --(-0.8,1.5) node[pos=0.8,left]{$t$};
    \end{tikzpicture}
    \caption{Complexified coordinates for the AdS black holes: the time coordinate
is complex in the extended spacetime, but with constant imaginary part in
each segment. The wedges are separated by the horizons
at $r = r_b$. The direction of increasing time in the right wedge is upwards and the direction of increasing time in left wedge is downwards. In the main text time in the left and right wedge is denoted by $t_L$ and $t_R$ respectively.}
    \label{fig:complextads}
\end{figure}

As a first example, consider the Schwarzschild black hole in Anti-de Sitter spacetime 
$SAdS_{d+1}$ in the infinite mass limit \cite{Fidkowski:2003nf}, for which the blackening factor reduces to
\begin{eqnarray}
    f(r)=r^2-\frac{1}{r^{d-2}} \, , 
\end{eqnarray}
where we introduced a dimensionless radial coordinate $r$. In this metric the event horizon sits at $r_b=1$ and the surface gravity is given by ${\kappa}=\frac{f'(r)}{2}\big|_{r=r_h}=\frac{d}{2}$, determining the inverse temperature as $\beta=1/T=2\pi/\kappa=4\pi/d$.
We define the tortoise coordinate as
\begin{eqnarray}\label{equ:rstar-SAdS}
r^*=\int_0^r\frac{dr'}{f(r')} + C\,,
\end{eqnarray}
where $C$ is an integration constant that we fix as $C=i\frac{\beta}{4}=\frac{i \pi}{d}$ to ensure that $r^*(r)$ is real everywhere outside the black hole horizon, i.e., for $r > r_b$. 
Identifying Kruskal coordinates 
\begin{eqnarray}
&&U \equiv T-X =  -e^{-{\kappa} ( t- r^*)} = -e^{-{\kappa} x^-}, \\
&&V \equiv T+X = e^{{\kappa}( t+ r^*)} = e^{{\kappa} x^+}, \label{X1}
\end{eqnarray}
the metric, ignoring the angular directions, is now regular at the horizon 
\begin{eqnarray}
    ds^2 = g(T,X)(-dT^2 + dX^2) \, .\label{krusg} 
\end{eqnarray}
Importantly, the Kruskal coordinates can now be globally extended from the region $(I)$ outside the black hole horizon, where both $t$ and $r^*$ are real and $V>0$ and $U<0$, to the other three regions in the $(U,V)$ plane where the signs of $U$ and $V$ are different (see figure \ref{fig:complextads}). In the other regions the static time coordinate $t$ then requires an imaginary part in order to agree with the corresponding signs of the Kruskal coordinates. For instance, $U$ and $V$ should both be positive in region $II$ inside the black hole horizon, implying that that Im$(x_+)=0$ and Im$(x_-)=-\beta/2$ (modulo integer multiples of $\beta$), which fixes Im$(t)=-\beta/4$ in that region. Applying the same logic one determines the imaginary parts of the static time coordinate in the other regions $III$ and $IV$ according to figure \ref{fig:complextads}. This procedure straightforwardly generalizes to other spacetimes with (multiple) horizons, as we will see.
Throughout the text, we will denote the real parts of these times as $\tau=\text{Re}(t)$, while their imaginary parts are determined by the corresponding region.

\begin{figure}[t]
\centering
\begin{subfigure}[b]{0.45\textwidth}
    \centering
\includegraphics[scale=1.0,trim=100 530 330 80,clip]{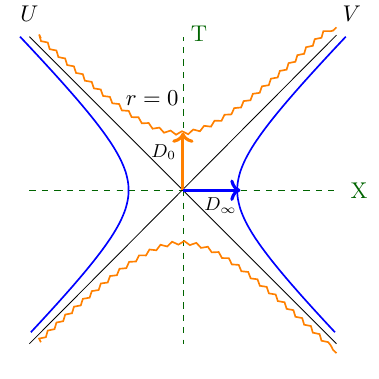}
    \caption{BTZ}
    \label{fig:btz}
\end{subfigure}
\hfill
\begin{subfigure}[b]{0.45\textwidth}
    \centering
\includegraphics[scale=1.0,trim=100 530 330 80,clip]{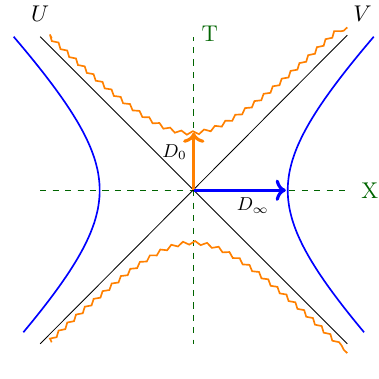}
    \caption{SAdS$_D$, with $D>3$}
    \label{fig:ads_d}
\end{subfigure}
\caption{Comparison of BTZ and SAdS$_D$ ($D>3$) Kruskal diagrams. In the former case $|D_0|=|D_\infty|$, while for the latter $|D_0|\leq|D_\infty|$. The orange lines represent the singularity $r=0$, and the blue ones correspond to the boundary.}
\label{fig:comparison-sads}
\end{figure}

In order to analyze the Kruskal diagram, we determine the curves of constant radius $r$, defined by $UV=-e^{2\kappa r^*}$ at the singularity $r=0$ and at the AdS boundary $r \rightarrow \infty$.
In dimensions $D=3, 4$ and $5$ one finds 
\begin{eqnarray}
UV = T^2 - X^2 &=& -e^{2\kappa r^*}\nonumber \\
&=& 
\begin{cases}
-\left(\frac{r-1}{r+1}\right) e^{2\arctan(r)} & \text{ } D=5\,, \\
-\frac{(r-1)}{\sqrt{r^2 + r + 1}} \exp\left(\sqrt{3} \arctan\left(\frac{2r+1}{\sqrt{3}}\right)-\frac{\pi}{2\sqrt{3}}\right) & \text{ } D=4\,, \\
-\left(\frac{r-1}{r+1}\right) & \text{ } D=3\,.
\end{cases}
\label{adsbh}
\end{eqnarray}
The resulting curves of constant $T^2 - X^2$ (see figure \ref{fig:comparison-sads}) show an important difference between $D=3$ (BTZ) and higher-dimensional AdS black holes: in the former case, the singularity and AdS boundary hyperbolic curves appear symmetrically, at the same coordinate distance from the origin in the Kruskal diagram.

\paragraph{Case 1:} For BTZ black holes, we set $d=2$ to find
\begin{align}    D_0&=\lim_{r\rightarrow0}UV=1\,,\nonumber\\
   D_\infty& =\lim_{r\rightarrow\infty}UV=-1\,.\label{x}
\end{align}
As a consequence the hyperbolic curves describing the singularity at $r=0$ and the AdS boundary at $r \rightarrow \infty$, since $|D_0| = |D_\infty|$, are rotated versions of each other, at a fixed distance from the origin (see figure \ref{fig:btz} for example). This is not the case for the $D>3$ AdS black hole geometries.

\paragraph{Case 2:} For $D=5$ AdS (large) black holes, one instead finds
\begin{eqnarray}
  D_0 &&= \lim_{r\rightarrow0}UV = 1\,,\nonumber\\
  D_\infty &&= \lim_{r\rightarrow\infty}UV = -e^{\pi}\,.
\end{eqnarray}
And similarly, for $AdS_4$ black holes, we have
\begin{eqnarray}
  D_0&& = \lim_{r\rightarrow0}UV = 1\,,\nonumber\\
  D_\infty &&= \lim_{r\rightarrow\infty}UV =-e^{\frac{\pi}{\sqrt{3}}}\,.
\end{eqnarray}
Note that for $D>3$ AdS black holes, $|D_0| < |D_\infty|$, which means that the singularity at $r=0$ is always closer to the origin than the AdS boundary. As a consequence the causal Penrose diagram is a perfect square for the SAdS geometry in $D=3$, but in higher dimensions the singularity bends towards the origin, relative to the AdS boundary.   
\begin{figure}[t]
    \centering
\begin{tikzpicture}[scale=1]
\draw[thick] (0,0) -- (0,4);
\draw[thick] (4,0) -- (4,4);
\draw[densely dotted] (0,0) -- (4.2,4.2);
\draw[densely dotted] (-0.2,4.2) -- (4,0);



\draw[thick, cyan] (0,2) -- (2,4);
\draw[thick, cyan] (4,2) -- (2,4);
\draw[thick, cyan] (0,2) -- (2,0);
\draw[thick, cyan] (4,2) -- (2,0);

\draw[dotted] (0,2) -- (4,2) node[pos=1.02,right]{$\tau_R=0$} node[pos=-0.02,left]{$\tau_L=0$};


\draw[->,thick] (2.7,1.5) .. controls (2.5,2) .. (2.7,2.5) node[pos=0.8,right]{$t$};
\draw[<-,thick] (1.3,1.5) .. controls (1.5,2) .. (1.3,2.5) node[pos=0.2,left]{$t$};

\tikzset{decoration={snake,amplitude=.15mm,segment length=2mm,
                       post length=.6mm,pre length=.6mm}};
\draw[orange,thick,decorate] (0,0) -- (4,0);
\draw[orange,thick,decorate] (0,4) -- (4,4);
\draw[black, line width=0.6mm, -{Latex[length=2.5mm, width=2.5mm]}] (4.2,4.2) -- node[above=1mm] {${V}$} (4.3,4.3);
\draw[black, line width=0.6mm, -{Latex[length=2.5mm, width=2.5mm]}] (-0.2,4.2) -- node[above=1mm] {${U}$} (-.3,4.3);\end{tikzpicture}
\caption{Penrose diagram for $D=3$ SAdS (BTZ).}
    \label{btz}
\end{figure}
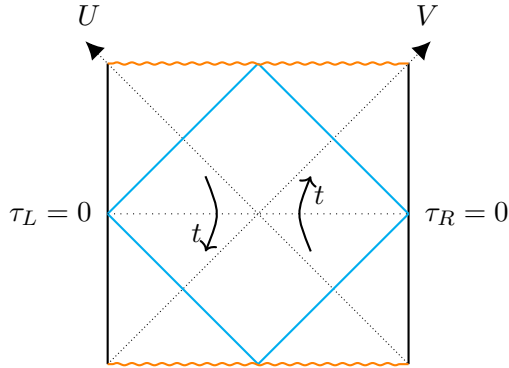

\begin{figure}[h!]
\centering
\begin{subfigure}[]{0.45\textwidth}
\centering
\begin{tikzpicture}[scale=1]
\draw[thick] (0,0) -- (0,4);
\draw[thick] (4,0) -- (4,4);
\draw[densely dotted] (0,0) -- (4.2,4.2);
\draw[densely dotted] (-0.2,4.2) -- (4,0);

\draw[thick, cyan] (2,3.4) -- (0,1.41);
\draw[thick, cyan] (2,3.4) -- (4,1.41);
\draw[thick, cyan] (2,0.59) -- (4,2.59);
\draw[thick, cyan] (2,0.59) -- (0,2.59);

 \draw[black, line width=0.6mm, -{Latex[length=2.5mm, width=2.5mm]}] (4.2,4.2) -- node[above=1mm] {${V}$} (4.3,4.3);
\draw[dotted] (0,2) -- (4,2);

\draw[green] (0,2) -- (1.36,3.45);
\draw[green] (4,2) -- (2.63,3.45);
\draw[green] (0,2) -- (1.36,0.55);
\draw[green] (4,2) -- (2.64,0.55);
\draw[black, line width=0.6mm, -{Latex[length=2.5mm, width=2.5mm]}] (-0.2,4.2) -- node[above=1mm] {${U}$} (-.3,4.3);
\filldraw[blue] (4,2.6) circle (2pt) node[anchor=west]{\scriptsize $\mathcal{T}$};
\filldraw[blue] (4,1.36) circle (2pt) node[anchor=west]{\scriptsize $-\mathcal{T}$};

\draw[thick, color=orange, decorate, decoration={snake,amplitude=.15mm,segment length=2mm, post length=.6mm,pre length=.6mm}] (0,4) to[out=-30, in=-150] (4,4);
\draw[thick, color=orange, decorate, decoration={snake,amplitude=.15mm,segment length=2mm, post length=.6mm,pre length=.6mm}] (0,0) to[out=30, in=150] (4,0);
\end{tikzpicture}
\caption{}
\label{fig:penrose-ax}
\end{subfigure}
\begin{subfigure}[]{0.45\textwidth}
\centering
\begin{tikzpicture}[scale=1]
\draw[densely dotted] (0,0) -- (4.2,4.2);
\draw[densely dotted] (-0.2,4.2) -- (4,0);

\draw[dotted] (-1,2) -- (5,2); 

\draw[thick, cyan] (0,2) -- (-0.85,1.07);
\draw[thick, cyan] (0,2) -- (-0.85,2.81);

\draw[thick, cyan] (4,2) -- (4.85,1.07);
\draw[thick, cyan] (4,2) -- (4.85,2.81);
\draw[thick, cyan] (0,2) -- (2,4);
\draw[thick, cyan] (4,2) -- (2,4);
\draw[thick, cyan] (0,2) -- (2,0);
\draw[thick, cyan] (4,2) -- (2,0);
\draw[thick, green] (-1,2) -- (1,4);
\draw[thick, green] (-1,2) -- (1,0);

\draw[thick, green] (5,2) -- (3,4);
\draw[thick, green] (5,2) -- (3,0);


\draw[->,thick] (2.7,1.5) .. controls (2.5,2) .. (2.7,2.5) node[pos=0.8,right]{$t$};
\draw[<-,thick] (1.3,1.5) .. controls (1.5,2) .. (1.3,2.5) node[pos=0.2,left]{$t$};
\draw[thick, color=black] (0,4) to[in=150, out=210] (0,0);

\draw[thick, color=black] (4,0) to[in=-30, out=30] (4,4);
 \draw[black, line width=0.6mm, -{Latex[length=2.5mm, width=2.5mm]}] (4.2,4.2) -- node[above=1mm] {${V}$} (4.3,4.3);
\filldraw[blue] (4.87,2.77) circle (2pt) node[anchor=west]{\scriptsize $\mathcal{T}$};

\filldraw[blue] (4.85,1.07) circle (2pt) node[anchor=west]{\scriptsize $-\mathcal{T}$};

\tikzset{decoration={snake,amplitude=.15mm,segment length=2mm, post length=.6mm,pre length=.6mm}};
\draw[orange,thick,decorate] (0,0) -- (4,0);
\draw[orange,thick,decorate] (0,4) -- (4,4);
\draw[black, line width=0.6mm, -{Latex[length=2.5mm, width=2.5mm]}] (-0.2,4.2) -- node[above=1mm] {${U}$} (-.3,4.3);
\end{tikzpicture}
\caption{}
\label{xyzchapa}
\end{subfigure}
\caption{Penrose diagrams of the SAdS$_D$ geometry with $D>3$. The dotted $45^{\circ}$ lines stand for the black hole horizon, whereas the blue ones stand for null rays. In figure (a), the singularity appears to bend relative to the straight vertical lines representing the asymptotic AdS boundary observers. In figure (b), the boundary is depicted as bowed outward relative to the horizontal straight line representing the singularity.
}
\label{fig:SAdspenrose}
\end{figure}
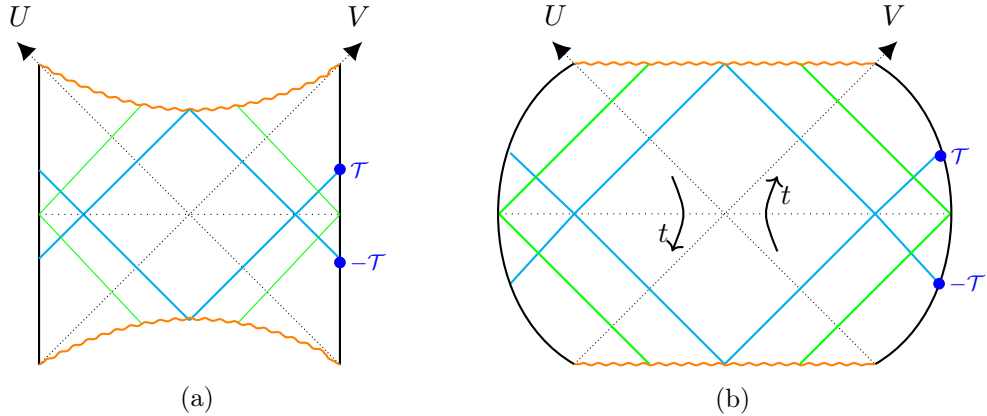

By inverting $X \rightarrow -X$ and $T \rightarrow -T$ the fully extended spacetime is bounded by four hyperbolic curves 
\begin{equation}
   D_\infty \leq T^2 - X^2 \leq D_0 \, . 
\end{equation}
From the Kruskal diagrams of (infinite mass) SAdS we can then easily construct the causal Penrose diagrams as shown in figure \ref{fig:penrose-ax}, where the black hole singularities bend inwards, coming closer to the origin of the diagram, relative to the AdS boundaries (represented as straight lines). Whenever $|D_0| < |D_\infty|$ radial null geodesics originating from the right boundary at $t = 0$ ($T = 0$) will reach the singularity ($X = 0$) right from the center. Only for $D=3$ is the Penrose diagram a perfect square, for $D>3$ the singularities, relative to straight AdS boundaries, bend inwards \cite{Fidkowski:2003nf}. Alternatively, if one decided to represent the black hole singularity by a straight line, the AdS boundaries would need to be bowed outwards, as illustrated in figure \ref{xyzchapa}, to appropriately describe the behavior of the radial null geodesics \cite{Fidkowski:2003nf,Faruk:2023uzs}. 

Radial null geodesics that are emitted from the right AdS boundary at \( \tau_R = 0 \), then reflect off the black hole future singularity, the other AdS boundary, and the past singularity, will only return to the exact same original AdS boundary point for $D=3$ (BTZ). For $D>3$ SAdS black holes, the reflected radial null geodesics end up at a later point \( \tau_R > 0 \), see figure \ref{fig:dsdxx}.  
Or equivalently, if one wishes to reflect a null ray from the center of the black hole singularity, the resulting symmetric null curve will connect two points on the conjugate AdS boundaries with opposite times (negative on the right, positive on the left), bounding the size of a symmetric causally connected wormhole region also known as a Wheeler-de Witt (WdW) patch, and defining the so-called critical time, as we will explain (and compute) in section \ref{xkutta}. As first pointed out in \cite{Fidkowski:2003nf} these reflected null curves probing the interior of the black hole geometry can be associated to (null) singularities in the analytically continued (symmetric) holographic CFT correlation function.
Moreover, as shown in \cite{Amado:2008hw}, reflected null rays probing the SAdS causal structure can be related to the asymptotic (eikonal) spectrum of quasinormal modes of the AdS black hole geometry. We will show that this interesting feature can be generalized to SdS geometries as well. But first let us remind the reader how to generalize the same kind of analysis to (Schwarzschild-) de Sitter spacetimes.



 \subsection{Black holes in de Sitter space}
Before discussing the analysis and results for the causal structure of Schwarzschild-de Sitter space, let us first briefly remind the reader of how to apply the same methods to empty de Sitter space. 

\paragraph{Pure de Sitter}

As is well-known, the causal structure of pure de Sitter space, summarized by the Penrose diagram, is a perfect square, implying that the causal diamonds of two observers maximally separated on the spatial spherer do not overlap and touch at a single (bifurcation) point. We can use exactly the same methods to quickly confirm this property. The blackening factor for empty dS spacetime, defining the metric in a single static patch region according to equation \eqref{general black}, is given by  
\begin{eqnarray}
    f(r)=1-\frac{r^2}{\ell^2} \, , 
\end{eqnarray}
where $\ell$ determines the cosmological event horizon, known as the de Sitter radius. This defines the surface gravity and the corresponding inverse de Sitter temperature as $\beta_c=2\pi/\kappa_c=2\pi \ell$. 
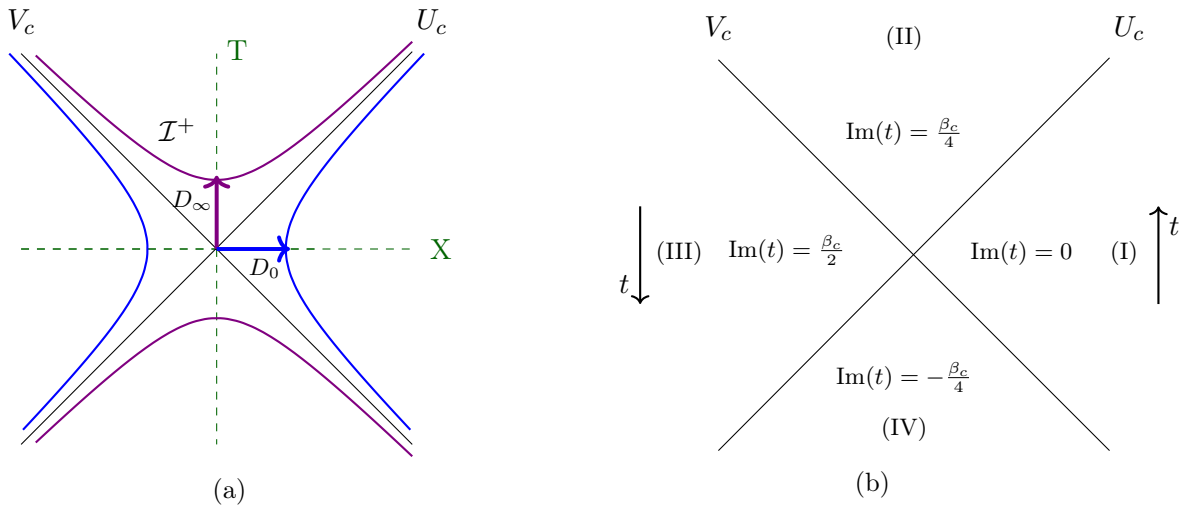
\begin{figure}[h!]
    \centering
    \begin{subfigure}[]{0.45\textwidth}
        \centering
        \begin{tikzpicture}[scale=1.3]
            \draw[mydarkgreen, dashed] (2,2) -- (2,4) node[pos=1.02,right] {T};
            \draw[mydarkgreen, dashed] (2,2) -- (2,0);
            \draw[mydarkgreen, dashed] (0,2) -- (4,2) node[pos=1.02,right] {X};
            \draw[mydarkgreen, dashed] (0,2) -- (3,2);
            \draw (0,0) -- (4,4.02);
            \draw (0,4) -- (4,0);
            \draw[violet, thick, domain=0.15:4, smooth, variable=\y, decorate, decoration={ amplitude=0.3mm, segment length=2mm}] plot ({\y}, {2 + sqrt(0.5 + (\y-2)^2)});
            \draw[violet, thick, domain=0.15:4, smooth, variable=\y, decorate, decoration={ amplitude=0.3mm, segment length=2mm}] plot ({\y}, {2 - sqrt(0.5 + (\y-2)^2)});
            \draw[blue, thick, domain=0.15:4, smooth, variable=\y] plot ({2 + sqrt(0.5 + (\y-2)^2)}, {\y});
            \draw[blue, thick, domain=0.15:4, smooth, variable=\y] plot ({2 - sqrt(0.5 + (\y-2)^2)}, {\y});
            \node[above] at (4.2,4.1) {$U_c$};
            \node[above] at (1.6,3.0) {$\mathcal{I}^+$};
            \draw[->, violet, line width=1.5pt] (2,2) -- (2,2.75);
            \draw[->, blue, line width=1.5pt] (2,2) -- (2.75,2);
            \node[above] at (0,4.1) {$V_c$};
            \node[above, font=\footnotesize] at (1.75,2.3) {$D_\infty$};
            \node[above, font=\footnotesize] at (2.49,1.6) {$D_0$};
        \end{tikzpicture}
        \caption{}
        \label{ccc-Dinf-D0}
    \end{subfigure}
    \hfill
    \begin{subfigure}[]{0.45\textwidth}
        \centering
        \begin{tikzpicture}[scale=1.3]
            \draw (0,0) -- (4,4.02);
            \draw (0,4) -- (4,0);
            \node[above] at (4.2,4.1) {$U_c$};
            \node[above] at (0,4.1) {$V_c$};
            \node[above, font=\footnotesize] at (1.9,3) {Im$(t)=\frac{\beta_c}{4}$};
            \node[above, font=\footnotesize] at (3.1,1.8) {Im$(t)=0$};
            \node[above, font=\footnotesize] at (1.9,0.5) {Im$(t)=-\frac{\beta_c}{4}$};
            \node[above, font=\footnotesize] at (1.9,4) {(II)};
            \node[above, font=\footnotesize] at (4.15,1.8) {(I)};
            \node[above, font=\footnotesize] at (-0.40,1.8) {(III)}; 
            \node[above, font=\footnotesize] at (1.9,0.0) {(IV)};
            \node[above, font=\footnotesize] at (0.7,1.8) {Im$(t)=\frac{\beta_c}{2}$}; 
            \draw[->,thick] (4.5,1.5) -- (4.5,2.5) node[pos=0.8,right]{$t$};
            \draw[->,thick]  (-0.8,2.5) --(-0.8,1.5) node[pos=0.8,left]{$t$};
        \end{tikzpicture}
        \caption{}
        \label{fig:complextsds}
    \end{subfigure}
     \caption{(a) Kruskal diagrams for empty dS$_D$ spacetime, where  $D_0$ and $D_\infty$ are given by $D_0 = \lim\limits_{r \rightarrow 0} U_c V_c$ and $D_\infty = \lim\limits_{r \rightarrow \infty} U_c V_c$. The two blue curves correspond to the north and south pole of the spatial $S^{(D-1)}$ sphere, mapping to the center of two conjugate static patches ($r=0$). The violet curves represent future and past spacelike infinity. (b) Complex static time of global dS space.}

\end{figure}

\begin{figure}[t]
\centering
\begin{tikzpicture}[scale=1.3]
\draw[thick,violet] (0,0) -- (4,0);
\draw[thick] (0,0) -- (0,4);
\draw[thick,violet] (0,4) -- (4,4);
\draw[thick] (4,0) -- (4,4);
\node[above] at (2,4) {$\mathcal{I}^+$};
\node[below] at (2,0) {$\mathcal{I}^-$};




\draw[dotted] (0,2) -- (4,2) node[pos=1.02,right]{$\tau_N=0$} node[pos=-0.02,left]{$\tau_S=0$};

\draw[densely dotted] (0,0) -- (4.2,4.2);
\draw[densely dotted] (-0.2,4.2) -- (4,0);
\draw[black, line width=0.6mm, -{Latex[length=2.5mm, width=2.5mm]}] (-0.2,4.2) -- node[above=1mm] {${V_c}$} (-.3,4.3);
 \draw[black, line width=0.6mm, -{Latex[length=2.5mm, width=2.5mm]}] (4.2,4.2) -- node[above=1mm] {${U_c}$} (4.3,4.3);
\end{tikzpicture}
\caption{Penrose diagram of de Sitter in arbitrary dimensions. 
}
\label{fig:SymmetricPenroseds}
\end{figure}
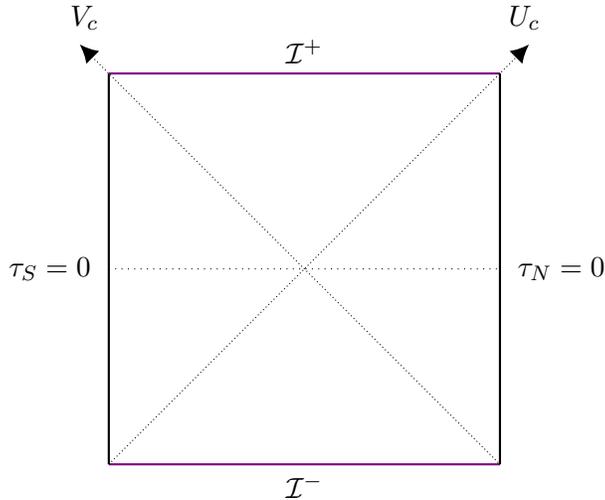

As before we will now extend the static coordinates beyond the horizon by introducing tortoise and then Kruskal type coordinates. Let's define the null coordinates $x^+$ and $x^-$ as
\be 
x^{\pm}=t\pm r^*\,,
\ee 
where $r^*(r)$ is the tortoise coordinate
\begin{eqnarray}
r^*=\int_0^r\frac{dr'}{f(r')}+C \, .
\end{eqnarray}
Here we pick the complex integration constant $C$ to vanish, to ensure that the tortoise coordinate is real inside the static patch region $r<\ell$. Note that for $r>\ell$ this convention implies that the tortoise coordinate acquires an imaginary part equal to $+\beta_c/4$.
This results in the following (dimension independent) solution for the tortoise coordinate in the static patch region $r<\ell$:
\begin{eqnarray}
    r^*=\frac{\ell}{2}\ln\left(\frac{l+r}{l-r}\right) . 
\end{eqnarray}
As before, we define Kruskal coordinates as follows, now labeled with a subscript $c$ to distinguish them from the black hole Kruskal coordinates 
\begin{eqnarray}
U_c&&=e^{\frac{1}{\ell}(t-r^*)}\,,\\
        V_c&&=-e^{-\frac{1}{\ell}(t+r^*)}\,,
\end{eqnarray}
which allows the de Sitter metric to be written as
\begin{eqnarray}
    ds^2= -\frac{4\ell^2}{(1-U_c V_c)^2} dU_c dV_c
+ \ell^2 \frac{(1+U_c V_c)^2}{(1-U_c V_c)^2}d\Omega^2_{d-2}\,.
\label{dsdim}
\end{eqnarray}
As the metric is now regular at the (past and future) horizon it can now be extended to the full $(U_c,V_c)$ plane, including two conjugate static patches associated to observers at the respective north or south pole of the spatial sphere $S^{D-1}$. Similar to the discussion in section~\ref{Kruskal coordinates of AdS blackhole spacetime}, when crossing the future horizon the static time coordinate now requires an imaginary contribution equal to $+i\beta_c/4$ to produce the correct signs {of the Kruskal coordinates}, see figure \ref{fig:complextsds}. This implies that the north and south poles, both associated to the center of the static coordinate patch $r=0$, are distinguished by the two different curves that solve $U_c V_c=-1$, whereas the (past and future) horizons correspond to $U_c V_c =0$ ($r=\ell$), and finally, the future and past spacelike infinity $\mathcal{I}^\pm$ curves are defined by $U_c V_c = 1$. As a result the Kruskal diagram is perfectly symmetric between the two observer and spacelike infinity curves, see figure \ref{ccc-Dinf-D0}. Transforming to the Penrose diagram, using spherical symmetry, one confirms the perfect square nature of de Sitter spacetime in arbitrary dimensions, see figure \ref{fig:SymmetricPenroseds}.


\paragraph{Schwarzschild-de Sitter}

Let us now introduce a second horizon in de Sitter spacetime, by adding a black hole. We will present a summarized discussion of the explicit construction in terms of Kruskal coordinates, for more details on null geodesics and the causal structure in $D=4,5$ and $6$ we refer to \cite{Faruk:2023uzs}. \\\\
For a black hole in de Sitter the blackening factor equals 
\begin{equation}
f(r) = 1 - \frac{r^2}{\ell^2} - \alpha\frac{M}{r^{D-3}}\label{bfactorSdS} \, .
\end{equation}
Here we introduced the constant $\alpha$, proportional to the gravitational constant $G$, equal to 
\begin{equation}
\alpha = \frac{16\pi G}{(D-2)\Omega_{D-2}}\,.
\end{equation}
In four dimensions this blackening factor can be factorized as follows 
 \begin{eqnarray} 
 f(r) = \frac{(r-r_b)(r_c-r)(r+r_b+r_c)}{\ell^2 r} \, , 
       \end{eqnarray}
in terms of the event horizon radii $r_b$ (for the black hole) and $r_c$ (for the cosmological horizon) expressed as \cite{Morvan:2022ybp}
 \begin{eqnarray} 
 r_{b,c} = r_N\left(\cos{\eta}\mp\sqrt{3}\sin{\eta}\right)  , 
       \end{eqnarray}
where we identified the parameter $\eta$
\begin{eqnarray}
    \quad\eta\equiv\frac{1}{3}\arccos{M/M_N}\, . 
\end{eqnarray}
Here we introduced the Nariai mass $M_N$, corresponding to the maximal mass that can fit in de Sitter spacetime. 
This maximal mass black hole is determined by the Nariai radius $r_N$, which equals 
\begin{equation}
r_N = \ell\sqrt{\frac{D-3}{D-1}}  \,,
\end{equation}
in dimensions $D>3$. In four dimensions this allows us to derive the following identity 
\begin{eqnarray}
r_b^2+r_c^2+r_br_c=3r_N^2\,.
\end{eqnarray}

As there exists both an interior black hole as well as an exterior cosmological horizon in the Schwarzschild-de Sitter (SdS) geometry, which confines observers at rest with respect to the black hole to a finite-size region, identifying the appropriate temperatures is ambiguous at best. A natural choice \cite{Morvan:2022ybp, Svesko:2022txo} is to identify the temperatures observed by the unique free-falling observer that does not cross any of the horizons, the so-called static sphere observer at \( r = r_\mathcal{O} \) \cite{Faruk:2023uzs}. The temperature detected by the static sphere observer at \( r = r_\mathcal{O} \) can be computed geometrically by determining the surface gravity at each of the two horizons (indicated by the subscript b or c)
\begin{equation}
    T_{b,c}^{(\mathcal{O})} = \frac{{\kappa}_{b,c}^{(\mathcal{O})}}{2\pi}\,.
\end{equation}
These temperatures will of course not be the same, except in the Nariai limit, implying that the black hole will decay to the empty maximal entropy de Sitter vacuum state. Strictly speaking of course, from a QFT point of view these temperatures are only well-defined under the assumption of an equilibrium. Formally however, we can define an interior black hole and exterior de Sitter equilibrium by imposing reflecting Dirichlet boundary conditions at the static sphere. This is the implicit assumption that we will use throughout this paper and it will have a couple of interesting implications. Physically it means that we will decouple the black hole interior from the de Sitter exterior, separated by a spherical mirror at the static sphere radius. Adding a scalar field consistent with these assumptions implies that the appropriate vacuum state is a direct product of an interior black hole and exterior de Sitter Hartle-Hawking effective thermo-field double construction, reproducing the (static sphere) interior and exterior temperatures. \\ 

As the unique radius where the cosmological expansion and the black hole attraction cancel each other, the static sphere radius $r_{\mathcal O}$ is determined by the stationary point of the blackening factor as
\begin{eqnarray}
    f'(r_{\mathcal{O}}) = 0\,.
\end{eqnarray}
In arbitrary dimensions the static sphere radius equals \cite{Morvan:2022ybp} (in terms of the black hole horizon \( r_b \)) 
\begin{eqnarray}
    r_{\mathcal{O}}^{D-1} = \frac{(D-3)}
    {2}r_{b}^{D-3}(\ell^2 - r_{b}^2)\,.
    \label{abdulKaku}
\end{eqnarray}
The standard definition of surface gravity,
\begin{equation}
    \zeta^\mu \nabla_\mu \zeta_\nu = \kappa \hs \zeta_\nu \,,
\end{equation}
introduces the Killing vector \( \zeta = \gamma^{-1} \partial_t \) 
with respect to each horizon, with \( \gamma \) being a normalization constant. As the blackening factor at the static sphere radius is \( f(r_{\mathcal{O}}) = 1 - (r_{\mathcal{O}}/r_N)^2 \), fixing the normalization to the proper time of the static sphere implies 
\begin{eqnarray}
    \gamma_\mathcal{O}=\sqrt{f(r_\mathcal{O})}= \sqrt{1 - (r_{\mathcal{O}}/r_N)^2} \, . 
    \label{SSnorm}
\end{eqnarray}
Having fixed the appropriate normalization we can compute the two surface gravities 
\begin{equation}
    {\kappa}_{b,c}^{(\mathcal{O})} = \frac{1}{2\gamma_\mathcal{O}}f'(r)\Big|_{r=r_{b,c}} \,,
\end{equation}
 of the black hole and cosmological horizons, respectively, with respect to the static sphere observer. In four dimensions this gives \cite{morvan:2022rn, Svesko:2022txo, Faruk:2023uzs}, where we introduced a superscript (and a subscript on $\gamma$ and $r$) to indicate that the temperatures correspond to those measured by observers at \( r = r_{\mathcal{O}} \)
\begin{eqnarray}
    {\kappa}_{b}^{(\mathcal{O})} &\equiv& \frac{\kappa_b}{\gamma_\mathcal{O}} = \frac{H^2}{2\gamma_\mathcal{O} r_b}(r_c - r_b)(r_c + 2r_b) \, , \label{100}\\
    {\kappa}_{c}^{(\mathcal{O})} &\equiv& \frac{\kappa_c}{\gamma_\mathcal{O}} = \frac{H^2}{2\gamma_\mathcal{O} r_c}(r_c - r_b)(2r_c + r_b)\,, \label{200}
\end{eqnarray}
where \( H \equiv \frac{1}{l} \) and the surface gravities $\kappa_{b/c}$ are those defined with the standard normalization $\gamma=1$. In the Nariai near-horizon limit the temperatures defined this way equal the temperature of the $2$-dimensional de Sitter spacetime \cite{Bousso:1996au}
\begin{eqnarray}
    {\kappa}_{N}^{(\mathcal{O})} = \frac{\sqrt{D-1}}{\ell}\,,
\end{eqnarray}
as one would expect. Instead, using the standard normalization (\(\gamma = 1\)) both temperatures would have vanished in the Nariai limit, even though the ($2$-dimensional de Sitter) entropy clearly remains finite. In the forthcoming sections, for the SdS geometry, we will always include the static sphere normalization factor $\gamma_\mathcal{O}$ when computing critical times.

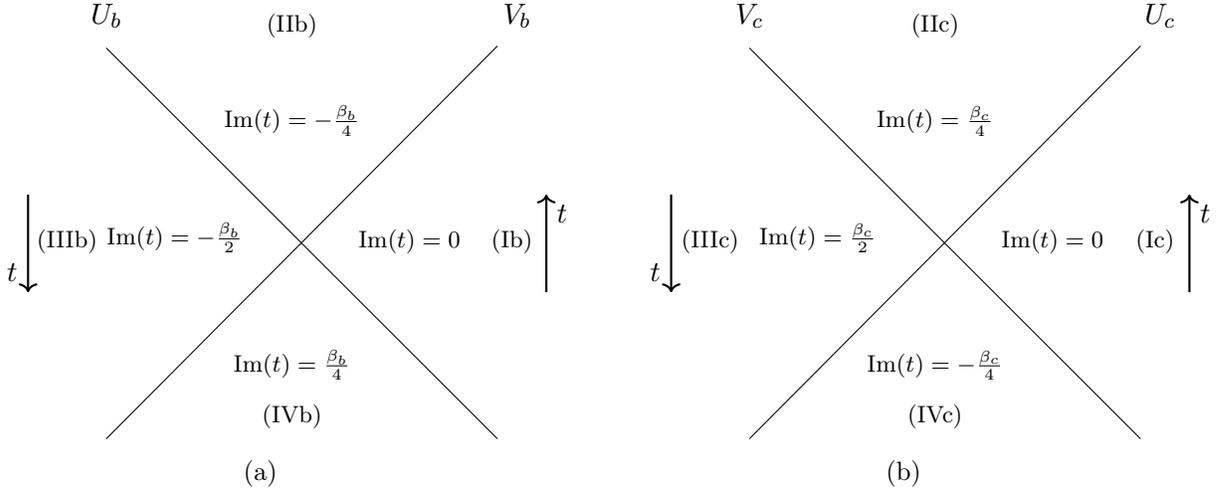
\begin{figure}[t]
    \centering
    \begin{subfigure}[b]{0.45\textwidth}
        \centering
        \begin{tikzpicture}[scale=1.3]
            \draw (0,0) -- (4,4.02);
            \draw (0,4) -- (4,0);
            \node[above, font=\footnotesize] at (1.9,4) {(IIb)};
            \node[above, font=\footnotesize] at (4.15,1.8) {(Ib)};
            \node[above, font=\footnotesize] at (-0.40,1.8) {(IIIb)}; 
            \node[above, font=\footnotesize] at (1.9,0.0) {(IVb)};
            \node[above] at (4.2,4.1) {$V_b$};
            \node[above] at (0,4.1) {$U_b$};
            \node[above, font=\footnotesize] at (1.9,3) {Im$(t)=-\frac{\beta_b}{4}$};
            \node[above, font=\footnotesize] at (3.1,1.8) {Im$(t)=0$};
            \node[above, font=\footnotesize] at (1.9,0.5) {Im$(t)=\frac{\beta_b}{4}$};
            \node[above, font=\footnotesize] at (0.7,1.8) {Im$(t)=-\frac{\beta_b}{2}$};   
            \draw[->,thick] (4.5,1.5) -- (4.5,2.5) node[pos=0.8,right]{$t$};
            \draw[->,thick] (-0.8,2.5) -- (-0.8,1.5) node[pos=0.8,left]{$t$};
        \end{tikzpicture}
        \caption{}
        \label{complextads}
    \end{subfigure}
    \hfill
    \begin{subfigure}[b]{0.45\textwidth}
        \centering
        \begin{tikzpicture}[scale=1.3]
               \draw (0,0) -- (4,4.02);
            \draw (0,4) -- (4,0);
            \node[above] at (4.2,4.1) {$U_c$};
            \node[above] at (0,4.1) {$V_c$};
            \node[above, font=\footnotesize] at (1.9,3) {Im$(t)=\frac{\beta_c}{4}$};
            \node[above, font=\footnotesize] at (3.1,1.8) {Im$(t)=0$};
            \node[above, font=\footnotesize] at (1.9,0.5) {Im$(t)=-\frac{\beta_c}{4}$};
            \node[above, font=\footnotesize] at (1.9,4) {(IIc)};
            \node[above, font=\footnotesize] at (4.15,1.8) {(Ic)};
            \node[above, font=\footnotesize] at (-0.40,1.8) {(IIIc)}; 
            \node[above, font=\footnotesize] at (1.9,0.0) {(IVc)};
            \node[above, font=\footnotesize] at (0.7,1.8) {Im$(t)=\frac{\beta_c}{2}$}; 
            \draw[->,thick] (4.5,1.5) -- (4.5,2.5) node[pos=0.8,right]{$t$};
            \draw[->,thick]  (-0.8,2.5) --(-0.8,1.5) node[pos=0.8,left]{$t$};
        \end{tikzpicture}
        \caption{}
        \label{ccc}
    \end{subfigure}
    \caption{ Complex time coordinates for the SdS geometry in (a) the interior black hole region, and (b) the exterior cosmological region. The black hole interior Kruskal coordinates $U_b$, $V_b$ cover the region $0<r<r_c$ while the exterior cosmological Kruskal coordinates cover the region $r_b<r<\infty$. 
   The domain Ib of the black hole region shares the same static sphere as the domain Ic of the cosmological region. Similarly, the domains IIIb and IIIc share the same static sphere, which is conjugate to the one shared by Ib and Ic.}
    \label{fig:imoft}
\end{figure}
We are now ready to introduce two sets of (extended) Kruskal coordinates for the SdS geometry. We will need to treat the extended interior black hole and extended exterior cosmological region separately, as a global cover of the geometry involving a single set of Kruskal coordinates clearly does not exist. 
We will start with the following definition of the tortoise coordinate in the region in between the black hole and cosmological horizon ($r_b<r<r_c$)
\begin{eqnarray}
r^* = \int_\infty^r \frac{dr'}{f(r')} + C \,.\label{tor}  
\end{eqnarray}
Choosing the integration constant as \(C = r^*(r=\infty)=\frac{i\pi}{2\kappa_c}\) ensures that \(r^*(r)\) is real in the region \(r_b < r < r_c\). In $D=4$ one can perform the integral to arrive at the following expression for the tortoise coordinate: 
\begin{eqnarray}
    r^*(r) =\frac{1}{2{\kappa}_c}\ln\left(\frac{r+r_b+r_c}{r_c - r}\right)-\frac{1}{2{\kappa}_b}\ln\left(\frac{r+r_b+r_c}{r - r_b}\right)  . 
    \label{tort}
\end{eqnarray}
Next, we first identify the appropriate Kruskal coordinates with respect to the black hole horizon
\begin{eqnarray}
U_b&&=
-e^{-{\kappa}_b ( t- r^*)} 
=-e^{-{\kappa}_b x^-}\,,\label{GHT}\\
V_b&&=e^{{\kappa}_b ( t+ r^*)}    =e^{{\kappa}_b x^+}\,.\label{X1b}
\end{eqnarray}
With this choice $U_b<0$ and $V_b>0$ in region Ib, the past and future horizon corresponds to $V_b=0$ and $U_b=0$ respectively, and the metric equals
\be 
ds^2= -\frac{f(r)}{\kappa_b^2}e^{-2{\kappa}_b r^*}dU_bdV_b+r^2d\Omega^2_{d-1}\,,
\ee 
where $r=r(U_b V_b)$. Recalling equation \eqref{tort}, it is straightforward to derive in four dimensions that
\be 
e^{-2{\kappa}_b r^*}=\left(\frac{r+r_b+r_c}{r - r_b}\right)\left(\frac{r_c - r}{r+r_b+r_c}\right)^{\kappa_b/\kappa_c},
\ee 
implying the following (extended) interior black hole metric in Kruskal coordinates 
\begin{eqnarray}
    ds^2=- \frac{(r_c-r)^{\frac{\kappa_b}{\kappa_c}+1}(r+r_b+r_c)^{-\frac{\kappa_b}{\kappa_c}+2}}{{\kappa_b^2 \ell^2 r}}
dU_bdV_b{+r^2d\Omega^2_{d-1}}\,.
    \label{toxr1}
\end{eqnarray}
This metric now covers the region $0<r<r_c$, as well as the conjugate interior black hole region, extending the static metric \eqref{bfactorSdS} beyond the black hole horizon.

To extend the metric beyond the cosmological horizon we identify a different set of Kruskal coordinates 
\begin{eqnarray}
U_c&&=e^{{\kappa}_c x^-}\,,\label{X2-c1}\\
V_c&&=-
e^{-{\kappa}_c x^+} \label{X2-c2}  \, , 
\end{eqnarray}
regular at $r=r_c$ with $U_c=0$ and $V_c=0$ at the past and future horizon respectively. For $r_b<r<r_c$ the Kruskal coordinates have now changed sign, with $U_c>0$ and $V_c<0$, as in the case of pure de Sitter, and can now be readily extended beyond the cosmological horizon  ($r_b<r<\infty$). 

In figure \ref{fig:imoft} we have illustrated how the static coordinates can be extended in the interior black hole and exterior cosmological region. Importantly, applying the same procedure as before, the static time coordinate acquires an imaginary contribution after crossing a horizon which is different for the interior black hole and exterior cosmological regions. As a consequence, in causal region III the imaginary value of the static time coordinate changes abruptly when crossing the (conjugate) static sphere, moving from the interior black hole IIIb to the exterior cosmological region IIIc. This is a direct consequence of the non-equilibrium nature of the SdS spacetime, preventing an everywhere regular Euclidean geometry\footnote{From a Euclidean geometry point of view we have selected periodicities that remove both conical singularities, implying a necessary shift in the Euclidean time circle somewhere in between, for instance at the static sphere \cite{morvan:2022rn}.}. As long as we are careful to assume Dirichlet (reflecting) boundary conditions at the static sphere, decoupling the interior black hole region from the exterior cosmological region, this discontinuity can be considered harmless at leading, semi-classical, order. Physically, this requires a reflecting physical object (with tension) at the static sphere, and we assume its backreaction can be neglected. Note that at the static sphere, semi-classically, this object will experience an excess (Hawking flux) pressure to expand into the exterior, which can be counteracted by internal pressures and/or by shifting its position slightly into the interior. As we fix the spacetime background, we will also ignore the interesting question of consistency of these Dirichlet boundary conditions in the full (backreacting) theory, which are known not to be elliptic in general~\cite{Witten:2018lgb,Anderson_2008,An:2021fcq,Anninos:2023epi,Anninos:2024wpy}. A more complete analysis of the stability and consistency of these assumed Dirichlet boundary conditions might certainly be of interest to study in more detail. 

The metric takes the form
\begin{eqnarray}
    ds^2=-\frac{f(r)}{{\kappa_c^2}}e^{2{\kappa}_c r^*(r)}dU_cdV_c{+r^2d\Omega^2_{d-1}}\,,
\end{eqnarray}
where $r=r(U_c V_c)$ and 
\begin{eqnarray}\label{eq-toget-Cc}
     e^{2{\kappa}_c r^*(r)} =\left(\frac{r+r_b+r_c}{r_c - r}\right)\left(\frac{r - r_b}{r+r_b+r_c}\right)^{\kappa_c/\kappa_b}.
\end{eqnarray}
As a consequence the extended metric on the exterior cosmological region is given by 
\begin{eqnarray}
    ds^2=-\frac{(r-r_b)^{\frac{{\kappa}_c}{{\kappa}_b}+1}(r+r_b+r_c)^{-\frac{{\kappa}_c}{{\kappa}_b}+2}}{{\kappa_c^2\ell^2 r }}dU_cdV_c +r^2d\Omega_{d-1}^2\,.
    \label{tor1}
\end{eqnarray}
\begin{figure}[t!]
\centering
\begin{subfigure}[]{0.45\textwidth}
    \centering
\includegraphics[scale=1.0,trim=100 530 330 80,clip]{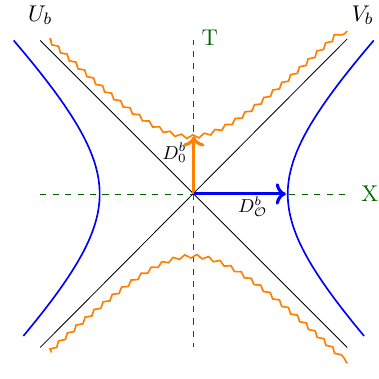} 
    \caption{Kruskal diagram of the interior black hole region of the four-dimensional SdS spacetime. The blue hyperbolas correspond to the static sphere and the orange hyperbolas represent the past and future singularity.}
    \label{fig:dsdxa}
\end{subfigure}
\hfill
\begin{subfigure}[]{0.45\textwidth}
    \centering
\includegraphics[scale=1.0,trim=100 530 330 80,clip]{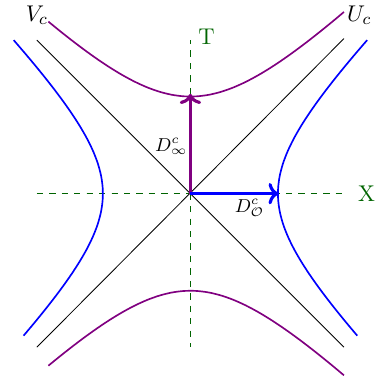}
    \caption{Kruskal diagram of the exterior cosmological region of the four-dimensional SdS spacetime.
    The blue hyperbolas correspond to the static sphere and the violet hyperbolas represent the past and future infinity $\mathcal{I}^\pm$.}
    \label{fig:dsdxb}
\end{subfigure}
\caption{Two Kruskal diagrams covering the full SdS spacetime.}
\label{fig:comparison-sds}
\end{figure}

\begin{figure}[t]
    \centering
    \begin{subfigure}[t]{0.48\linewidth}
        \centering
        \includegraphics[width=\linewidth]{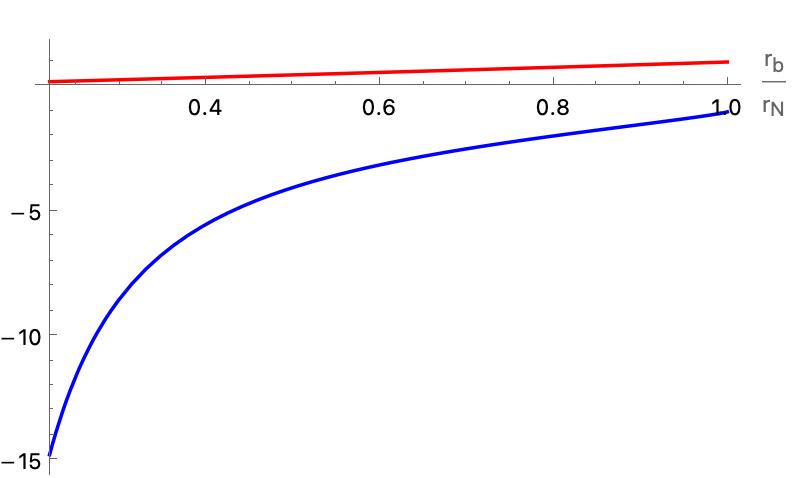}
        \caption{Plot of the quantities $D^b_0$ (red) and $D^b_{\mathcal{O}}$ (blue).}
        \label{jrocX1662}
    \end{subfigure}%
    \hfill
    \begin{subfigure}[t]{0.48\linewidth}
        \centering
        \includegraphics[width=\linewidth]{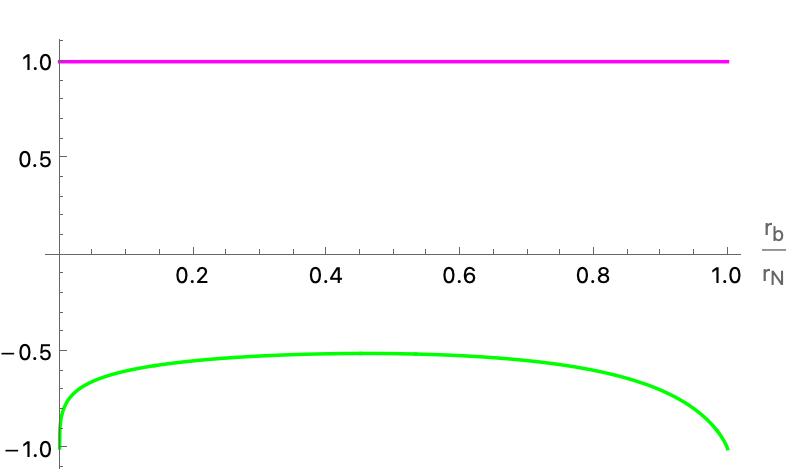}
        \caption{Plot of $D^c_\infty$ (magenta) and $D^c_\mathcal{O}$ (green).}
        \label{kashukhon}
    \end{subfigure}
    \caption{The distance of each hyperbola from the center in the Kruskal diagrams for both SdS regions: (a) the interior black hole region, and (b) the exterior cosmological region.
}
    \label{fig:combined_plots}
\end{figure}

We are now ready to analyze and plot the (causal) structure of both the interior black hole and the exterior cosmological region. For the interior black hole region, to construct the Kruskal diagram of figure \ref{fig:dsdxa}, we use equations (\ref{GHT}) and (\ref{X1b}) to write  
\begin{eqnarray}
    U_b V_b=-e^{2{\kappa}_b r^*}.
       \label{UbVb}
\end{eqnarray}
The above equation identifies a hyperbolic curve of fixed $X^2- T^2$, labeled by a constant $r$ radius sphere in the interior black hole region. In the (extended) Kruskal diagram covering the interior black hole region, the shortest coordinate distance to the origin of the two hyperbolic curves representing the singularity $r=0$ and the static sphere $r=r_{\mathcal{O}}$ is defined as 
\begin{subequations}
\begin{eqnarray}
   D^b_0 &=& \lim_{r\rightarrow0}U_bV_b=-\lim_{r\rightarrow0}e^{2{\kappa}_b r^*}\,, \label{distance1} 
   \\
   D^b_{\mathcal{O}} &=& \lim_{r\rightarrow r_{\mathcal{O}}}U_bV_b=
       -\lim_{r\rightarrow r_{\mathcal{O}}}       
       e^{2{\kappa}_b r^*}. \label{distance2} 
\end{eqnarray}
\end{subequations}
Computing this explicitly in $D=4$ for any mass parameter $M$, see figure \ref{jrocX1662}, we conclude that
\begin{align}
|D^b_0|\leq|D^b_{\mathcal{O}}| \, ,
\label{xxxx1}
\end{align} 
where the equality sign corresponds to the Nariai limit. So we confirm that the black hole singularity bends inward, relative to the static sphere observer. 

Let us now analyze the exterior cosmological region of the SdS geometry, covered by the extended Kruskal coordinates $U_c$ and $V_c$ in equations \eqref{X2-c1} and  \eqref{X2-c2}. The shortest coordinate distance to the origin of the two hyperbolas representing (past or future) spacelike infinity and the static sphere are defined as 
\begin{subequations}
\begin{eqnarray}
   D^c_\infty &=& \lim_{r\rightarrow\infty}U_cV_c=-\lim_{r\rightarrow\infty}e^{-2{\kappa}_c r^*}\,, \label{distance3} 
   \\
   D^c_{\mathcal{O}} &=& \lim_{r\rightarrow r_{\mathcal{O}}} U_cV_c=
       -\lim_{r\rightarrow r_{\mathcal{O}}}       
       e^{-2{\kappa}_c r^*} \,.\label{distance4} 
\end{eqnarray}
\end{subequations}
\begin{figure}[t]
 \centering
    \resizebox{0.7\linewidth}{!}{
        \begin{tikzpicture}[scale=0.9] 
              \node at (2.25,3.9) {\scriptsize $r=0$};
            \node[above] at (6.7,4.8) {\scriptsize $\mathcal{I}^+$}; 
            \node[above] at (6.7,-0.8) {\scriptsize $\mathcal{I}^-$}; 
              \node at (2.25,0.6) {\scriptsize $r=0$};
              \node[above] at (5.3,5.0) {$V_b$};
               \node[above] at (3.80,4.3) {$U_c$};
                      \node[above] at (9.4,5.0) {$V_c$};
                      \node[above] at (0,4.80) {$U_b$};
\draw[green] (4.5,0) -- (6.75,2.25);
\draw[green] (4.5,0) -- (2.25,2.25);
\draw[green] (4.5,4.5) -- (2.25,2.25);

            \draw[cyan,thick]  (0,1.36) -- (2.25,3.63);
            \draw[cyan,thick]  (0,3.18) -- (2.25,.88);
            \draw[cyan,thick]  (4.5,3.18) -- (2.25,.88);
            \draw[cyan,thick]  (4.5,1.35) -- (2.25,3.63);
            \draw[red,thick]  (4.5,2.59) -- (6.75,4.84);
            \draw[red,thick]  (9,2.59) -- (6.75,4.84);
            \draw[red,thick]  (9,1.93) -- (6.75,-.34);
            \draw[red,thick]  (4.5,1.93) -- (6.75,-.34);

           \draw[dashed] (9,0) -- (9,4.5);

            \draw[dashed] (0,0) -- (0,4.5);
            \draw[thick] (4.5,0) -- (4.5,4.5);
            \draw[densely dotted] (0,0) -- (2.25,2.25);
            \draw[densely dotted] (9,0) -- (6.75,2.25);
            \draw[densely dotted] (6.75,2.25) -- (9,4.5);
            \draw[densely dotted] (2.25,2.25) -- (0,4.5);
            \node[above] at (0,-0.7) {\scriptsize $r=r_{\mathcal{\mathcal{O}}}$}; 
            \node[above] at (4.5,-0.7) {\scriptsize $r=r_{\mathcal{\mathcal{O}}}$}; 
            \node[above] at (9,-0.7) {\scriptsize $r=r_{\mathcal{\mathcal{O}}}$}; 
            \draw[thick, color=orange, decoration={snake,amplitude=.17mm,segment length=1.5mm, post length=.5mm}, decorate] (0,4.5) to[out=-40, in=-140] (4.5,4.5);
            \draw[thick, color=orange, decoration={snake,amplitude=.17mm,segment length=1.5mm, post length=.5mm}, decorate] (0,0) to[out=40, in=140] (4.5,0);
            \draw[dotted] (0,2.25) -- (9,2.25);
            \draw[thick, color=violet] (4.5,4.5) to[out=15, in=165] (9,4.5);
            \draw[thick, color=violet] (4.5,0) to[out=-15, in=-165] (9,0);
              \filldraw[red] (4.5,2.6) circle (2pt) node[anchor=west]{\scriptsize ${\tau_{\mathcal{O}}}=\mathcal{T}_c$};
              \filldraw[red] (4.5,1.9) circle (2pt) node[anchor=west]{\scriptsize ${\tau_{\mathcal{O}}}=-\mathcal{T}_c$};
            \draw[densely dotted] (4.5,4.5) -- (4.3,4.7);
            \draw[densely dotted] (4.5,4.5) -- (4.7,4.7);
            \filldraw[cyan] (4.5,3.15) circle (2pt) node[anchor=east]{\scriptsize ${\tau_{\mathcal{O}}}=\mathcal{T}_b$};

              \filldraw[cyan] (4.5,1.35) circle (2pt) node[anchor=east]{\scriptsize ${\tau_{\mathcal{O}}}=-\mathcal{T}_b$};

\draw[ green] (4.5,4.53) -- (6.75,2.25);
\draw[black, line width=0.6mm, -{Latex[length=2.5mm, width=2.5mm]}] (-0.2,4.7) -- (-0.4,4.9);

\draw[black, line width=0.6mm, -{Latex[length=2.5mm, width=2.5mm]}] (4.3,4.7) -- (4.1,4.9);\draw[black, line width=0.6mm, -{Latex[length=2.5mm, width=2.5mm]}] (4.7,4.7) -- (4.9,4.9);

\draw[black, line width=0.6mm, -{Latex[length=2.5mm, width=2.5mm]}] (4.3,4.7) -- (4.1,4.9);\draw[black, line width=0.6mm, -{Latex[length=2.5mm, width=2.5mm]}] (9.2,4.7) -- (9.4,4.9);
            \draw[densely dotted] (0,4.5) -- (-.2,4.7);

            \draw[densely dotted] (9,4.5) -- (9.2,4.7);

        \end{tikzpicture}
    }

\caption{Four-dimensional SdS Penrose diagram. As a convention, we refer to the causal region in the middle of the interior black hole region as the right static patch, bounded by the right static sphere (vertical bold line). The conjugate causal region is referred to as the left static patch, bounded by the left static sphere (vertical dashed line). Similarly, the right static sphere bounds the exterior de Sitter region in the middle, while the left static sphere bounds the conjugate exterior de Sitter region.  
This Penrose diagram illustrates our earlier choice in figure \ref{ccc}, where domains $I_b$ and $I_c$ share the same static sphere.}  
\label{fig:SdSabdul}
\end{figure}
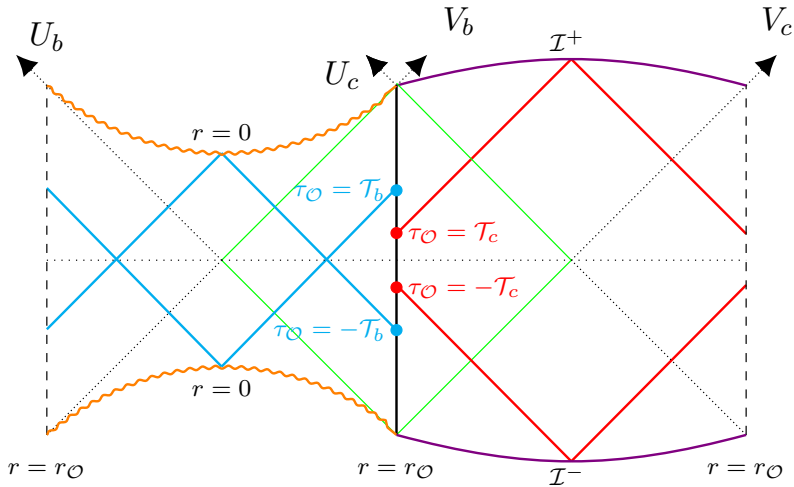

In figure \ref{kashukhon} we plotted the results, and comparing the two distances we conclude  
\begin{eqnarray}
|D^c_\infty|\geq|D^c_{\mathcal{O}}| \, , 
\label{xxxx2} 
\end{eqnarray}
where the equality sign corresponds to the empty de Sitter and Nariai limit. Combining the results of the interior black hole and exterior cosmological region we conclude that with respect to the static sphere, the black hole singularity bends inward and that cosmological spacelike infinity bends outwards, only to equal (two) perfect squares in the Nariai limit. The corresponding Kruskal and Penrose diagrams for four-dimensional SdS are found in figures \ref{fig:comparison-sds} and \ref{fig:SdSabdul}, but have a similar structure in $D=5$ and $D=6$ \cite{Faruk:2023uzs}.  
This particular structure implies that, for a generic mass of the black hole, no ingoing radial null geodesic originating from the static sphere at \(\tau_{\mathcal{O}} = 0\) (\(T = 0\)) reaches the center of the singularity (\(X = 0\)) in the Penrose diagram. Instead a null geodesic has to be emitted at \(\tau_{\mathcal{O}} = -\mathcal{T}_b\) to reach the center of the singularity \cite{Faruk:2023uzs}. 
Similarly, for the asymptotic cosmological region, we observe that outgoing radial null geodesics emitted from the static sphere reach the center of future spacelike infinity when released at \(\tau_{\mathcal{O}} = \mathcal{T}_c\). The special times \(\mathcal{T}_b\) and \(\mathcal{T}_c\) depend on the mass of the black hole, but reduce to the symmetric point $\mathcal{T}_{b/c}=0$ in the limit of empty de Sitter (dS) and the Nariai limit \cite{Faruk:2023uzs}. 

Having introduced the methods and basic results for the causal structure of AdS and dS black hole geometries, we will now proceed to discuss a couple of interesting consequences and applications. We will first introduce a purely geometric derivation of the asymptotic quasinormal mode spectrum for the SdS geometry, and then move on to introduce the (additional) effect of shock wave perturbations on the interior and exterior Wheeler-de Witt (WdW) causal regions, determined by the critical times.

\newpage

\section{Null geodesics and the asymptotic quasinormal mode spectrum}

We will first briefly review eikonal, large frequency, techniques for computing the s-wave ($l=0$) quasinormal modes (QNMs) for black hole horizons. For the details, we refer to \cite{Fidkowski:2003nf} and \cite{Amado:2008hw}. For more background information on quasinormal modes in general, their potential relevance in the partition function, and to compare interesting limits of our final results with existing literature, we refer the reader to some classic reviews and results \cite{Horowitz:1999jd, Denef:2009kn, Konoplya:2011qq, Berti:2009kk, Konoplya:2004uk, Festuccia:2008zx, Cardoso:2001hn, Lopez-Ortega:2006aal, Du:2004jt, Abdalla:2002hg}, as well as more recent works \cite{Konoplya:2022xid,Konoplya:2024ptj,Sarkar:2023rhp,Chrysostomou:2023jiv, Law:2023ohq}.   
The essential point which allows one to relate radial null geodesics connecting conjugate points to quasinormal modes, is that in the geodesic approximation the correlator between heavy field operators evaluated at these conjugate events, separated by an event horizon, can be mapped to the thermal correlator of field operators in the same causal region, with one of the operators evaluated at a complex time. To be more specific, introducing a large mass scalar field operator $\phi$ and a vacuum state $\ket{\Psi}$ that effectively produces a thermofield double state in a single causal patch, we can write the thermal correlator in that region (labeled as region R) as the expectation value
\begin{equation}
\bra{\Psi} \phi_R(0) \phi_R(t)  \ket{\Psi} = \mathrm{Tr} \, \rho_\beta \, \phi_R(0) \phi_R(t)  \, . 
\end{equation}
Using the thermofield double construction we can also define correlators involving both a right (R) and left (L) causal region, for example-
\begin{equation}
\bra{\Psi} \phi_R(0) \phi_L(t)  \ket{\Psi}  \, ,
\end{equation}
which, as we saw, can be computed by analytically continuing time to complex values in the field operator in the right region and is therefore contained in the thermal correlator
\begin{equation}
\bra{\Psi} \phi_R(0) \phi_R(-t-i \beta/2)  \ket{\Psi} = \mathrm{Tr} \, \rho_\beta \, \phi_R(0) \phi_R(-t-i\beta/2)  \, .
\end{equation}
This implies that the analytically continued thermal correlator, in a geodesic approximation, is naively expected to exhibit a lightcone singularity when the two (conjugate) points are connected by a null geodesic. This identifies a particular value of the real time difference, fixed by the details of the causal structure, and with the imaginary part equal to $\beta/2$, moving one point into the conjugate causal region. As it turns out the relevant geodesics that contribute in the thermofield double state are complex, preventing the appearance of a lightcone singularity. But a careful analysis of the late-time analytic behavior of these complex geodesics does reproduce the asymptotic quasinormal mode frequencies \cite{Fidkowski:2003nf}, which can be derived purely geometrically by studying the eikonal approximation and null geodesics in the relevant geometry with horizons \cite{Amado:2008hw}. This means that information about the detailed causal structure of the geometry, extended beyond the horizon, is probed by correlation functions in a (single) causal region, outside of the horizon.

As a reminder, the eikonal approximation is a high-frequency limit where $\omega \gg R$, with $R$ representing the typical curvature scale of the spacetime. In the eikonal approximation the scalar field is written as 
\begin{eqnarray}
    \phi(u, v) = A(u) e^{i \omega \, v}\,,
\end{eqnarray}
and the field equations are then expanded in the large frequency limit, with $A(u) \sim v \sim \mathcal{O}(\omega^0)$. In this limit, classical solutions to the equations of motion of a quantum field theory can be described using the eikonal approximation. The  coordinates $(u,v)$ are known as Rosen coordinates, where $u$ is aligned with the affine parameter along a null geodesic and the phase $v$ satisfies the (leading order) eikonal equation 
\begin{eqnarray}
    g^{\mu\nu} \partial_\mu v \, \partial_\nu v = 0 \, . 
\end{eqnarray}
This implies that the parameter $v$ defines a congruence of null rays whose tangent vector is null, which can also be understood as the Hamilton-Jacobi function of the (geodesic) variational problem. As a consequence the propagation of field perturbations in the bulk spacetime in the large frequency limit is well approximated by null rays at constant eikonal phase $\omega \, v$, reducing the problem to one of geometric optics in curved spacetime.

For a spherically symmetric spacetime described by equation~\eqref{general black}, the conserved quantity $\mathcal{E}$ associated with null radial geodesics is determined as\footnote{The derivative is with respect to the affine parameter $\lambda$.} \cite{Fidkowski:2003nf,Faruk:2023uzs}
\begin{eqnarray}
    \dot{r} = \pm\mathcal{E} \label{xdiskee} \, .
\end{eqnarray}
Fixing the observer at a constant radius $r = r_\mathcal{O}$, defining the starting point of the null geodesic, depending on the signs we can distinguish four different solutions. For $\mathcal{E} > 0$ the two solutions are (where we choose $u$ 
to vanish at the starting point $r = r_\mathcal{O}$) 
\begin{eqnarray}
&&    \dot{r} = -\mathcal{E}, \quad u = -(r-r_\mathcal{O}), \quad v = -t - r^* \label{1}, \\
&&    \dot{r} = \mathcal{E}, \quad u = (r-r_\mathcal{O}), \quad v = -t + r^* \label{2}.
\end{eqnarray}
And for $\mathcal{E} < 0$
\begin{eqnarray}
&&    \dot{r} = -\mathcal{E}, \quad u = (r-r_\mathcal{O}), \quad v = t + r^* \label{3}, \\
&&    \dot{r} = \mathcal{E}, \quad u = -(r-r_\mathcal{O}), \quad v = t - r^* \label{4}.
\end{eqnarray}
In these expressions $r^* = \int \frac{dr}{f(r)}$ is the usual tortoise coordinate.  

For a specific geometry under consideration, we need to identify the conserved quantities and endpoints of the radial null geodesics. We will first quickly remind the reader of the results of Amado and Hoyos \cite{Amado:2008hw}, who identified a constant eikonal phase while crossing the horizon to compute the asymptotic spectrum of s-wave scalar field quasinormal modes for Schwarzschild black holes in AdS spacetime. After having reviewed this analysis for AdS black holes, we will then extend these ideas to Schwarzschild-de Sitter (SdS) geometries and discuss the resulting asymptotic spectrum of s-wave quasinormal modes associated with both the cosmological and black hole horizons. 

\subsection{Asymptotic quasinormal mode frequencies for AdS black holes}

\begin{figure}[t!]
    \centering
\includegraphics[scale=1.0,trim=100 530 330 80,clip]{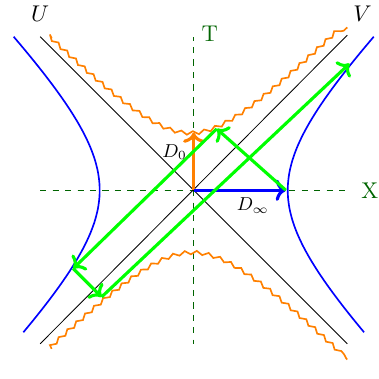}
    \caption{For the $SAdS_D$ geometry in $D>3$, a null geodesic starts from the right boundary at $\tau_R=0$, bounces off the future singularity, reaches the left AdS boundary, bounces off the past singularity, and finally returns to the boundary at $\tau_R' > \tau_R=0$. }
    \label{fig:dsdxx}
\end{figure}

The Schwarzschild-AdS (SAdS) black hole geometry has a singularity at $r = 0$ 
and a timelike boundary at $r \rightarrow \infty$,
where we assume the observer is located, as depicted in the Penrose diagram (see figure \ref{fig:SAdspenrose}). We will start by identifying and then connecting the null geodesics in the SAdS geometry, using equations~\eqref{1}–\eqref{4}. Note that the constant $v$ rays defined by equations \eqref{1} and \eqref{2} differ from equations \eqref{3} and \eqref{4} by just the sign of $t$, implying that decreasing/increasing radial null geodesics moving forward in time in region I connect to increasing/decreasing radial null geodesics moving forward in time in the conjugate region III. Equivalently, radial null geodesics starting at the right boundary moving inwards towards the future singularity connect (after reflection) to radial null geodesics moving outwards towards the left boundary. By identifying which null rays connect at the (future) singularity one can derive a complex static time difference at the AdS boundary between the starting- and end-point of the radial null geodesic. This complex time difference can then be associated to a (null) singularity pole in the Keldysh-Schwinger correlation function, which can be related to quasinormal mode frequencies after a Fourier transformation.  

We want to derive this complex time difference for an infinite mass black hole in a $D$-dimensional AdS geometry. Due to the inward bending of the future and past singularities, for null geodesics starting at one of the AdS boundaries, then bouncing off the singularity and reaching the conjugate AdS boundary, one can easily derive that the (complex) static time shift equals 
\begin{eqnarray}
    \Delta t=\Delta v|_{r^*=0}= 2\int_0 ^\infty \frac{dr}{f(r)}=\int_{-\infty} ^\infty \frac{du}{f(u)}\,,\label{chummadetankk}
\end{eqnarray}
{where we extended the domain of $f$ to negative values as $f(u)=f(-u)$ for $u<0$.}
The result for SAdS is straightforward to compute and gives 
\begin{eqnarray}
     \Delta t = 2\mathcal{T}_b - i\frac{\beta}{2} \, , \label{juakheli}
\end{eqnarray}
with the imaginary part arising from crossing two horizons and the real part $\mathcal{T}_b$, corresponding to the so-called critical time as we will explain later, equal to 
\begin{eqnarray}
    \mathcal{T}_b = \frac{\beta}{4} \cot{\left(\frac{\pi}{d}\right)} \, . \label{T}
\end{eqnarray}
Notably, this complex time shift computes which points at the two separated AdS boundaries are connected by a (reflected) null geodesic and as a consequence suggests the appearance of a lightcone singularity in the Schwinger-Keldysh field theory correlator in the eikonal, large frequency, limit. The asymptotic s-wave ($l=0$) spectrum of quasinormal modes is then straightforwardly derived from equation \eqref{juakheli} by performing a Fourier transform \cite{Amado:2008hw, Fidkowski:2003nf,Festuccia:2008zx}
\begin{eqnarray}
    \omega_n \approx n\pi\frac{\mathcal{T}_b \pm i\frac{\beta}{4}}{\mathcal{T}_b^2+(\frac{\beta}{4})^2}, \label{jk}
\end{eqnarray}
For BTZ black holes, where the past and future singularity do not bend inwards, the real part of the complex time difference vanishes, telling us that the QNM frequencies are purely imaginary. The particular sign of the imaginary part depends on the convention for the Fourier transform (and the scalar field ansatz), but should result in a characteristic exponential decay of the perturbation \cite{Cardoso:2001hn,Amado:2008hw,Fidkowski:2003nf}. 

In general this approach teaches us that the real and imaginary parts of the s-wave QNM spectrum, in the large frequency limit, are directly related to the extended causal structure of the geometry, see figure \ref{fig:dsdxx}. The imaginary part, resulting in exponential decay, is given by the horizon temperature and the real part will be non-vanishing in dimensions $D>3$, where the singularities bend inwards, leading to oscillatory behavior. An important (implicit) condition for this connection between the causal structure and the (asymptotic) QNM spectrum is the presence of Dirichlet boundary conditions for the (massive) scalar field at the AdS boundary and purely ingoing boundary conditions at the black hole horizon. As we will explain next, we can generalize this structure (and the corresponding boundary conditions) to black holes in de Sitter space and derive an asymptotic QNM spectrum for both the interior black hole and exterior de Sitter region, that can be associated to observers at the static sphere, using the same geometric approach. 

\subsection{Asymptotic quasinormal mode frequencies in SdS geometries}

To study the asymptotic QNM spectrum in SdS geometry we will follow exactly the same procedure. In the SAdS geometry the null geodesics started from the AdS boundary, where one assumes the natural (CFT) observers to be located. In the absence of a boundary for the SdS geometry the natural place to identify as a starting point for null geodesics (and accompanying boundary conditions for a massive scalar field) is the static sphere. Indeed, in the two limits of vanishing black hole mass or vanishing cosmological constant, the static sphere radius reduces to the static patch center and spatial infinity respectively. It is also the natural location for imposing reflecting Dirichlet boundary conditions, separating the interior black hole region from the exterior de Sitter region, as we already emphasized earlier. This allows one to consistently select an interior black hole and exterior de Sitter state by removing the conical singularity in imaginary time, producing an interior and exterior thermofield double state. These are exactly the ingredients we need in order to follow the same recipe for deriving an interior and exterior set of asymptotic s-wave QNM frequencies using just the properties of (reflecting) radial null rays. Of course, it is straightforward to relate the complex time differences computed for the static sphere to other fixed radii, for example the stretched horizons.  

So let us proceed and compute the interior black hole and exterior de Sitter complex time shifts as measured by an observer on the static sphere, using equation \eqref{abdulKaku}. In the split SdS geometry the two different directions
of future evolving radial null geodesic can be characterized as follows: one moving inward towards the black hole future singularity and the other moving outward towards de Sitter future infinity. We will use the superscripts $b$ and $c$ to distinguish between these radial null geodesics.
The causal structure of the interior and exterior regions is different and described by equations~(\ref{xxxx1}) and (\ref{xxxx2}) respectively. Considering the black hole region in \(D > 3\) dimensional SdS spacetime, a null geodesic starting from the right static sphere at \(\tau_{\mathcal{O}} = 0\) bounces off the future black hole singularity to reach the left static sphere. 
In contrast, focusing on the cosmological region, an outgoing radial null ray emitted from the right static sphere at \(\tau_{\mathcal{O}} = 0\) bounces off future infinity \(\mathcal{I}^+\) to arrive at the left static sphere.
The qualitative differences in the behavior of interior and exterior radial null geodesics translate to different properties of the asymptotic QNM spectrum associated to the interior black hole and exterior de Sitter region.

To continue, for both the interior black hole and exterior de Sitter region we need to normalize the complex time difference to the proper time of the static sphere radius by introducing the appropriate redshift factor $\gamma_\mathcal{O}$ from equation \eqref{SSnorm}. The SdS geometry features an interior past and future black hole singularity and an exterior past and future de Sitter infinity, as portrayed in the Penrose diagram (see figure \ref{fig:SdSabdul}), and the straight bold and dashed vertical lines correspond to the static sphere, with the left and right static sphere boundaries identified.  
Now consider radial null geodesics in the interior black hole region of the SdS geometry.  We can start from equations~\eqref{1}–\eqref{4} and just rescale the coordinate time to the proper time at $r=r_\mathcal{O}$.
The two solutions for $\mathcal{E} > 0$ are
\begin{eqnarray}
&&    \dot{r} = -\mathcal{E}, \quad u = -(r-r_\mathcal{O}), \quad v = -t^b -  r^* \label{.1}, \\
&&    \dot{r} = \mathcal{E}, \quad u = r-r_\mathcal{O}, \quad v = -t^b +  r^* \label{.2}.
\end{eqnarray}
As before, for $\mathcal{E} < 0$, we find the other two solutions for radial null geodesics in the black hole interior region
\begin{eqnarray}
&&    \dot{r} = -\mathcal{E}, \quad u = r-r_\mathcal{O}, \quad v =   t^b +  r^* \label{.3}, \\
&&    \dot{r} = \mathcal{E}, \quad u = -(r-r_\mathcal{O}), \quad v = t^b -  r^* \label{.4}.
\end{eqnarray}
\begin{figure}[t!]
\centering
\begin{subfigure}[]{0.45\textwidth}
    \centering
\includegraphics[scale=1.0,trim=100 530 330 80,clip]{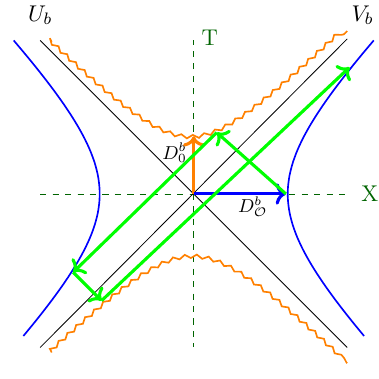}
    \caption{ 
}
    \label{fig:dsdxa-null}
\end{subfigure}
\hfill
\begin{subfigure}[]{0.45\textwidth}
    \centering
\includegraphics[scale=1.0,trim=100 530 330 80,clip]{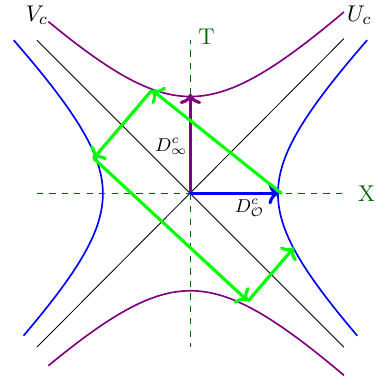}
    \caption{}
    \label{fig:dsdxb-null}
\end{subfigure}
\caption{Bouncing null geodesics in the (a) interior
and (b) exterior of SdS geometry. }
\label{fig:comparison-sds-null}
\end{figure}

As before, the null geodesic described by equation \eqref{.1} starts from the right static sphere and proceeds towards the future singularity at $r=0$. Upon reflection, equation \eqref{.2} identifies the radial null geodesic starting from the future singularity ending up at the left static sphere. 
The (complex) time shift, as measured by a static sphere observer and defining the location of a lightcone singularity in the scalar field correlator, is defined in this case as 
\begin{eqnarray}
    \Delta t^{b}_{\mathcal{O}}=\gamma_{\mathcal{O}} \Delta t^b=2\gamma_{\mathcal{O}}
    \int_0^{r_{\mathcal{O}}}
    \frac{dr}{f(r)} \, ,
\end{eqnarray}
where we introduced $\gamma_{\mathcal{O}}$ to give the proper time of the static sphere observer. 

Similarly, for the null geodesics bouncing around in the exterior dS region, we find two solutions for $\mathcal{E} > 0$ 
\begin{eqnarray}
&&    \dot{r} = \mathcal{E}, \quad u = (r-r_{\mathcal{O}}), \quad v = - t^c +  r^*\label{..1}\,, \\
&&    \dot{r} = -\mathcal{E}, \quad u = -(r-r_{\mathcal{O}}), \quad v = - t^c -  r^*  \label{..2}\,.
\end{eqnarray}
And for $\mathcal{E} < 0$, 
\begin{eqnarray}
&&    \dot{r} = \mathcal{E}, \quad u = -(r-r_{\mathcal{O}}), \quad v =  t^c -  r^* \label{..3}\,, \\
&&    \dot{r} = -\mathcal{E}, \quad u = (r-r_{\mathcal{O}}), \quad v =   t^c +   r^* \label{..4}\,.
\end{eqnarray}
Again, equation~\eqref{..1} describes radial null geodesics that start at the right static sphere and move towards de Sitter future infinity. Bouncing off de Sitter future infinity, equation~\eqref{..2} describes radial null geodesics that start at de Sitter future infinity and travel towards the left static sphere. 
As a consequence the complex time difference for two events connected by an exterior radial null geodesic emitted from the right static sphere, crossing two cosmological horizons and ending up at the left static sphere, can be determined performing the following integral 
\begin{eqnarray}
    \Delta t^c_{\mathcal{O}}=\gamma_\mathcal{O}\Delta t^c=2\gamma_\mathcal{O}
    \int_{r_{\mathcal{O}}}^\infty
    \frac{dr}{f(r)}\,.
\end{eqnarray}
With all the ingredients in place we can now compute for the SdS geometry, first for the black hole interior region,  the complex time shift as measured by an observer on the static sphere
\begin{eqnarray}
     \Delta t^b_{\mathcal{O}} = 2\mathcal{T}_b^{\mathcal{O}} - i\frac{\beta_b^{\mathcal{O}}}{2} \, ,  \label{juakhelib}
\end{eqnarray}
as well as for the de Sitter exterior region
\begin{eqnarray}
     \Delta t^c_{\mathcal{O}} = -2\mathcal{T}_c^{\mathcal{O}} + i\frac{\beta_c^{\mathcal{O}}}{2}\,.\label{juakheli-c}
\end{eqnarray}
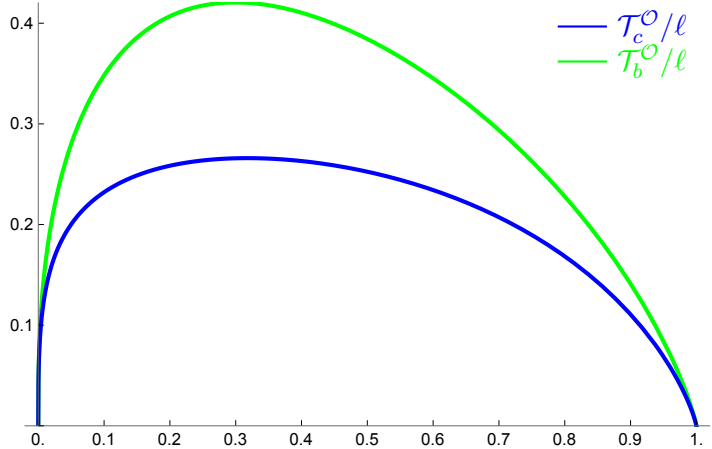
\begin{figure}[]
  \centering
  \begin{tikzpicture}
    \node[anchor=south west, inner sep=0] (image) at (0,0) 
      {\includegraphics[width=0.6\textwidth]{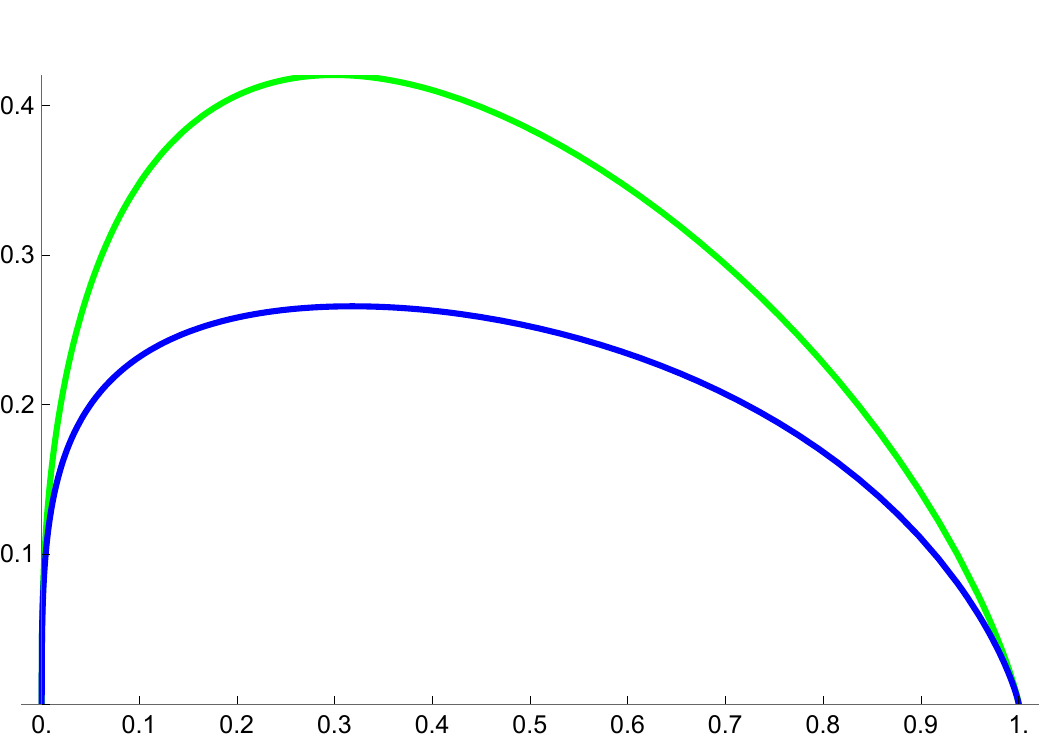}};
    \begin{scope}[x={(image.south east)},y={(image.north west)}]
      \draw[blue, thick] (0.78, 0.85) -- (0.85, 0.85); 
      \node[anchor=west, blue] at (0.85, 0.85) {$\mathcal{T}_c^{\mathcal{O}}/\ell$};
      
      \draw[green, thick] (0.78, 0.78) -- (0.85, 0.78); 
      \node[anchor=west, green] at (0.85, 0.78) {$\mathcal{T}_b^{\mathcal{O}}/\ell$};
    \end{scope}
  \end{tikzpicture}
  \caption{Normalized critical times $\mathcal{T}_c^{\mathcal{O}}/\ell$ (blue) and $\mathcal{T}_b^{\mathcal{O}}/\ell$ (green) as a function of the black hole radius ${r_b}/{r_N}$ in four dimensions.
  }
\end{figure}
The real parts of the complex time shift in $\Delta t^b$ or $\Delta t^c$ are determined by the details of the causal structure in the interior black hole and exterior de Sitter regions of the SdS geometry respectively. In the interior black hole region a radial null geodesic generically bounces off the future singularity off-center, to the right, whereas in the exterior de Sitter region it generically reflects at future dS infinity off-center to the left. From there, a Fourier transformation results in the complex s-wave asymptotic QNM spectrum associated to the interior black hole and exterior de Sitter region respectively. Starting from equations \eqref{juakhelib} and \eqref{juakheli-c} the final results for the complex QNM frequencies are 
\begin{eqnarray}
    \omega_n^{b,\mathcal{O}}\approx n\pi\frac{\mathcal{T}_b^{\mathcal{O}}\pm i\frac{\beta_{b}^{\mathcal{O}}}{4}}{(\mathcal{T}_b^{\mathcal{O}})^2+(\frac{\beta_b^{\mathcal{O}}}{4})^2}\,, \label{jkchuu}\\
       \omega_n^{c,\mathcal{O}} \approx n\pi\frac{\mathcal{T}_c^{\mathcal{O}}\pm i\frac{\beta_c^{\mathcal{O}}}{4}}{(\mathcal{T}_c^{\mathcal{O}})^2+(\frac{\beta_c^{\mathcal{O}}}{4})^2} \, , \label{jk-c}
\end{eqnarray}
where the signs of the imaginary parts are fixed by the conventions for the time evolution and Fourier transform, resulting in the characteristic exponential decay for both the interior black hole and exterior de Sitter region. 
For the four-dimensional SdS geometry one can derive the following analytic expressions for the critical times $\mathcal{T}_b^{\mathcal{O}}$ and $\mathcal{T}_c^{\mathcal{O}}$ that determine the real parts of the complex time differences
\begin{eqnarray}
 &&   \mathcal{T}_b^{\mathcal{O}}=\frac{1}{2\kappa^{\mathcal{O}}_c}\ln\left(\frac{(r_{\mathcal{O}}+r_b+r_c)r_c}{(r_c-r_{\mathcal{O}})(r_b+r_c)}\right)
    -
    \frac{1}{2\kappa^{\mathcal{O}}_b}\ln\left(\frac{(r_{\mathcal{O}}+r_b+r_c)r_b}{(r_{\mathcal{O}}-r_{b})(r_b+r_c)}\right),\label{mapus1}\\
 &&   \mathcal{T}_c^{\mathcal{O}}=
   \frac{1}{2{\kappa}_c^{\mathcal{O}}}\ln\left(\frac{r_{\mathcal{O}}+r_b+r_c}{r_c - r_{\mathcal{O}}}\right)-\frac{1}{2{\kappa}_b^{\mathcal{O}}}\ln\left(\frac{r_{\mathcal{O}}+r_b+r_c}{r_{\mathcal{O}} - r_b}\right).\label{mapus2}
\end{eqnarray}

As it should, the real parts $\mathcal{T}_b^{\mathcal{O}}$ and $\mathcal{T}_c^{\mathcal{O}}$ are functions of the black hole mass. Just like BTZ black holes, in $D=3$ dimensions $\mathcal{T}_b^{\mathcal{O}}$ and $\mathcal{T}_c^{\mathcal{O}}$ vanish. For the SdS geometry in $D>3$, the real parts of the complex frequency are non-zero for a generic mass of the black hole, except in the special limiting cases of pure de Sitter and the Nariai black hole. This makes sense: in both limits the causal structure reduces to a perfect square, and quasinormal modes no longer oscillate, they just decay.  
As one should expect, in the \( r_b \rightarrow 0 \) pure de Sitter limit, equation \eqref{jk-c} reproduces the QNM behavior for any \( D \)-dimensional de Sitter spacetime, with \( \beta_{c}^{\mathcal{O}} \rightarrow \beta_{dS} = {2\pi \ell}\) \cite{VerlindeH:2023SYKdS,Lopez-Ortega:2006aal,Du:2004jt,Abdalla:2002hg,Chrysostomou:2023jiv,Law:2023ohq}. 
In the Nariai limit, when \( r_b = r_c \), we instead find \cite{Bousso:1996au,morvan:2022rn,Faruk:2023uzs,Eune:2012mv}
\begin{eqnarray}
\beta_{c}^{\mathcal{O}} = \beta_{b}^{\mathcal{O}} \rightarrow \beta_{N} = {2\pi \ell_N} \, ,      
\end{eqnarray} 
which indeed corresponds to the inverse Nariai temperature in terms of the Nariai radius \( \ell_N \), which is related to the dS radius \( \ell \) as follows
\begin{eqnarray}
    \ell_N = \frac{\ell}{\sqrt{D-1}}\,.
\end{eqnarray}
We note that the static sphere normalization plays a crucial role in reproducing the correct QNM frequencies in the Nariai limit. In that limit the geometry reduces to the direct product $dS_2\times S^{D-2}$ \cite{Bousso:1996au,Svesko:2022txo,morvan:2022rn}, with the black hole and cosmological horizons now in in thermal equilibrium at a temperature $T_N=\frac{1}{\beta_N}$. The asymptotic s-wave QNM frequencies in the exterior de Sitter region in equation (\ref{jk-c}) in these two different limits then indeed reproduces \cite{VerlindeH:2023SYKdS,Lopez-Ortega:2006aal,Du:2004jt,Abdalla:2002hg,Chrysostomou:2023jiv,Law:2023ohq}
\begin{eqnarray}
\omega_n^{\scalebox{0.6}{dS}}&& = \pm i \, 4\pi n\, T_{\scalebox{0.6}{dS}}\,,\\
\omega_n^{\scalebox{0.6}{Nariai}}&& = \pm i \, 4\pi n \, T_{\scalebox{0.6}{Nariai}}\,.
\end{eqnarray}
Again, the signs should be fixed to obtain exponentially decaying behavior in time.  
For a generic mass of the black hole, and in \( D > 3 \), the real parts, expressed in terms of the critical times, do not vanish and reflect the fact that the asymptotic QNM spectra in the s-wave sector, as measured by a static sphere observer, are oscillating at frequencies that are sensitive to the interior black hole and the exterior de Sitter causal structure respectively. As these results are only valid in the asymptotic, large $n$ limit, corresponding to large frequencies and decay factors, these modes are subdominant as compared to the leading $n=1$ response at late times. Nevertheless, the large $n$ modes might be useful probes of the partition function \cite{Denef:2009kn} and contain interesting information about a potentially holographic dual description \cite{Susskind:2021esx, VerlindeH:2023SYKdS, Verlinde:2024znh}.

\section{Critical times, shock waves and complexity in SdS spacetimes}

In the previous section we showed how the details of the causal structure of the SdS geometry determine the asymptotic QNM spectrum for the interior black hole and exterior de Sitter parts. To compute the complex time difference \cite{Fidkowski:2003nf} between two conjugate events connected by a reflected null geodesic, and the related frequency, we introduced the so-called critical time parameters. Here we will first remind the reader of another important application and interpretation of these critical time parameters. They identify the limiting, maximal, causal Einstein-Rosen bridge wormhole region connecting the two conjugate static patches \cite{Aalsma:2020aib,Galante:2023uyf}, which is also referred to as the Wheeler-de Witt (WdW) patch, whose volume or action \cite{Stanford:2014jda,Couch:2016exn, Baiguera:2023tpt} is conjectured to be related to the notion of quantum complexity, corresponding to a holographic probe of the time evolving geometry beyond the event horizon. At the critical time this quantity in fact starts to diverge, but by introducing appropriate cut-offs the late-time growth of complexity can be shown to be linear in $t$ \cite{Brown:2015lvg,Goto:2018iay,Anegawa:2023dad,Chapman:2021jbh}, consistent with the characteristic behavior of a chaotic quantum system. Before hitting the critical time the macroscopic complexity variable evolves relatively slowly, which is referred to as the `plateau' phase \cite{Chapman:2018dem,Brown:2015lvg}. Our main result is that we will explicitly compute the change in the critical times after introducing shock waves into the interior and exterior parts of the SdS geometry in $D>3$ to determine the SdS switchback time delays or advances \cite{Baiguera:2024xju, Anegawa:2023dad, Aguilar-Gutierrez:2023pnn}, from the natural perspective of the static sphere observer. We compute this for arbitrary black hole mass and at the static sphere, where the change in the critical time is affected by both the (well-known) local shift from the shock wave, as well as the deformed causal structure, generalizing earlier results \cite{Anegawa:2023dad,Chapman:2021jbh}. 
Assuming a holographic dual description the computed time delays should be related to the typical response to perturbations in quantum chaotic systems, tied to their scrambling behavior \cite{Shenker:2013pqa,Hayden:2007cs,Sekino:2008he}.

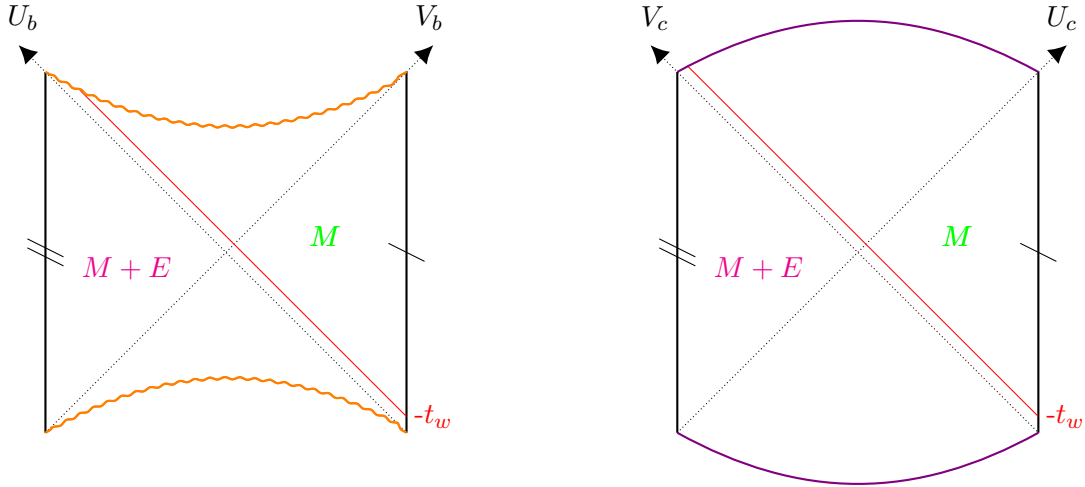
\begin{figure}[h!]
\centering
\begin{tikzpicture}[scale=1.2]
\draw[thick] (0,0) -- (0,4);
\draw[thick] (4,0) -- (4,4);

\draw[densely dotted] (0,0) -- (4.2,4.2);
\draw[densely dotted] (-0.2,4.2) -- (4,0);




\draw[black, line width=0.6mm, -{Latex[length=2.5mm, width=2.5mm]}] (4.2,4.2) -- node[above=1mm] {${V_b}$} (4.3,4.3);
\draw[red] (4,0.18)--(0.38,3.8);
\draw[black, line width=0.6mm, -{Latex[length=2.5mm, width=2.5mm]}] (-0.2,4.2) -- node[above=1mm] {${U_b}$} (-.3,4.3);



\node[red] at (4.3, 0.18) {-$t_w$};

\draw (3.8,2.1)--(4.2,1.9);
\draw (7+3.8,2.1)--(7+4.2,1.9);

\draw (3.8-4,2.15)--(4.2-4,1.95);
\draw (7+3.8-4,2.15)--(7+4.2-4,1.95);
\draw (3.8-4,2.05)--(4.2-4,1.85);
\draw (7+3.8-4,2.05)--(7+4.2-4,1.85);





\draw[thick, color=orange, decorate, decoration={snake,amplitude=.15mm,segment length=2mm, post length=.6mm,pre length=.6mm}] (0,4) to[out=-30, in=-150] (4,4);
\draw[thick, color=orange, decorate, decoration={snake,amplitude=.15mm,segment length=2mm, post length=.6mm,pre length=.6mm}] (0,0) to[out=30, in=150] (4,0);


\draw[thick] (7,0) -- (7,4);
\draw[thick] (11,0) -- (11,4);

\draw[densely dotted] (7,0) -- (11.2,4.2);
\draw[densely dotted] (7-0.2,4.2) -- (11,0);




\draw[black, line width=0.6mm, -{Latex[length=2.5mm, width=2.5mm]}] (11.2,4.2) -- node[above=1mm] {${U_c}$} (11.32,4.3);
\draw[red] (11,0.18)--(7.11,11.18-7.11);
\draw[black, line width=0.6mm, -{Latex[length=2.5mm, width=2.5mm]}] (7-0.2,4.2) -- node[above=1mm] {${V_c}$} (7-.3,4.3);



\node[red] at (11.3, 0.18) {-$t_w$};

\node[green] at (3.1, 2.18) {$M$};
\node[green] at (3.1+7, 2.18) {$M$};

\node[green] at (3.1, 2.18) {$M$};
\node[green] at (3.1+7, 2.18) {$M$};

\node[magenta] at (0.9, 2-.18) {$M+E$};
\node[magenta] at (0.9+7, 2-.18) {$M+E$};





\draw[thick, color=orange, decorate, decoration={snake,amplitude=.15mm,segment length=2mm, post length=.6mm,pre length=.6mm}] (0,4) to[out=-30, in=-150] (4,4);
\draw[thick, color=orange, decorate, decoration={snake,amplitude=.15mm,segment length=2mm, post length=.6mm,pre length=.6mm}] (0,0) to[out=30, in=150] (4,0);

            \draw[thick, color=violet] (11,4) to[out=151, in=29] (7,4);
            \draw[thick, color=violet] (11,0) to[out=-151, in=-29] (7,0);
\end{tikzpicture}
\caption{Penrose diagram of SdS deformed by a pair of shock waves emitted at the same time $t = -t_w$: one with positive energy $+E$ towards the asymptotically dS region and another with negative energy $-E$ into the black hole region. Causal regions I and III are identified, yielding a consistent causal structure by gluing the geometries at the shock waves. 
}
\label{fig:SdS-setup}
\end{figure}

An important observation is that the only (globally) consistent way to add radial shock waves to the SdS geometry is to add them in pairs with a total energy that vanishes \cite{Aalsma:2021kle}. This can be easily understood by realizing that locally gluing two SdS geometries with slightly different masses will run into inconsistencies matching these geometries along the compact periodic spatial slices, which can be fixed by introducing another (opposite energy, anti-) shock wave to return to the original SdS geometry \cite{morvan:2022rn,Faruk:2023uzs}, see figure \ref{fig:SdS-setup}. So in particular we will consider emitting a positive energy shock wave through the cosmological horizon, matched by a negative energy shock wave passing through the black hole horizon. This choice is motivated by the expectation that this will 'open up' both the interior black hole and exterior de Sitter wormhole region, and in addition is related to the natural evolution towards the maximum entropy pure de Sitter state\footnote{Of course, in our set-up where the interior black hole and exterior de Sitter region are decoupled, one can also choose to emit a positive energy shock wave into the black hole and a negative energy shock wave through the cosmological horizon.}. This (local) energy conserving consistency condition effectively couples the interior black hole and exterior de Sitter region in a controlled way. This is an important distinction from asymptotically AdS backgrounds, where one can freely change the boundary conditions to insert a shock wave \cite{Hirano:2019ugo,Hotta:1992wb}. To start, as before, we provide a reminder and quick review of these ideas and the relevant calculations for AdS black holes. 

\subsection{Critical time for black holes in AdS}
\label{xkutta}

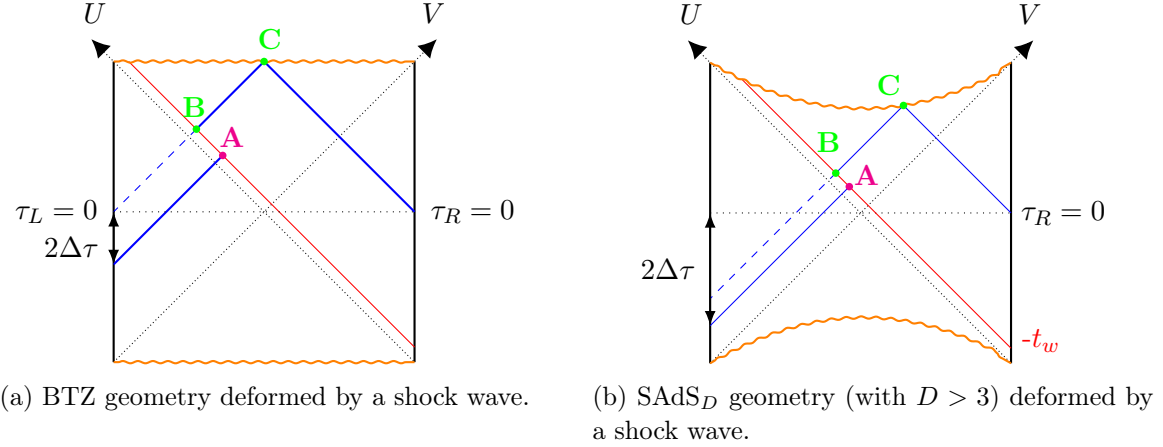
\begin{figure}[t]
    \centering
    \begin{subfigure}[t]{0.48\textwidth}
        \centering
        \begin{tikzpicture}[scale=1]
            \draw[thick] (0,0) -- (0,4);
            \draw[thick] (4,0) -- (4,4);
            \draw[densely dotted] (0,0) -- (4.2,4.2);
            \draw[densely dotted] (-0.2,4.2) -- (4,0);
            \draw[red] (4,0.2)--(0.2,4);
            
            \draw[thick, blue] (1.1,3.1) -- (2,4);
            \draw[dashed, blue] (1.1,3.1) -- (0,2);
            \draw[thick, blue] (1.45,2.75) -- (0,1.3);
            \draw[thick, blue] (4,2) -- (2,4);

            \tikzset{decoration={snake,amplitude=.15mm,segment length=2mm,
                               post length=.6mm,pre length=.6mm}};
            \draw[orange,thick,decorate] (0,0) -- (4,0);
            \draw[orange,thick,decorate] (0,4) -- (4,4);

            \draw[black, line width=0.6mm, -{Latex[length=2.5mm, width=2.5mm]}] (-0.2,4.2) -- node[above=1mm] {${U}$} (-.3,4.3);
            \draw[black, line width=0.6mm, -{Latex[length=2.5mm, width=2.5mm]}] (4.2,4.2) -- node[above=1mm] {${V}$} (4.3,4.3);

            \draw[dotted] (0,2) -- (4,2) node[pos=-0.02,left]{$\tau_L=0$} node[pos=1.02,right]{$\tau_R=0$};

            \draw[black, line width=0.2mm, <->, >=Latex] (0,1.3) -- 
            node[left=0.6mm, yshift=-3.5pt] {$2\Delta \tau $} 
            (0,2);

            \fill[green] (2,4) circle (0.05);
            \fill[green] (1.1,3.1) circle (0.05);
            \fill[magenta] (1.45,2.75) circle (0.05);
            \node[green] at (2, 4.3) {\textbf{ C}};
            \node[green] at (1.0, 3.4) {\textbf{ B}};
            \node[magenta] at (1.5, 3.0) {\textbf{ A}};
        \end{tikzpicture}
        \caption{BTZ geometry deformed by a shock wave.}
        \label{fig:btz-shock}
    \end{subfigure}
    \hfill
    \begin{subfigure}[t]{0.48\textwidth}
        \centering
        \begin{tikzpicture}[scale=1]
            \draw[thick] (0,0) -- (0,4);
            \draw[thick] (4,0) -- (4,4);
            \draw[densely dotted] (0,0) -- (4.2,4.2);
            \draw[densely dotted] (-0.2,4.2) -- (4,0);

            \draw[dotted] (0,2) -- (4,2);
            \draw (4,2) node[right]{$\tau_R=0$};

            \draw[blue] (4,2) -- (2.57,3.43);
            \draw[blue] (1.67,2.53) -- (2.57,3.43);
            \draw[blue,dashed] (1.67,2.53) -- (0,0.86);

            \draw[blue] (0,0.5) -- (1.85,2.35);
 \draw[red] (4,0.2)--(0.42,3.78);
            \draw[thick, color=orange, decorate, decoration={snake,amplitude=.15mm,segment length=2mm, post length=.6mm,pre length=.6mm}] (0,4) to[out=-30, in=-150] (4,4);
            \draw[thick, color=orange, decorate, decoration={snake,amplitude=.15mm,segment length=2mm, post length=.6mm,pre length=.6mm}] (0,0) to[out=30, in=150] (4,0);

            \draw[black, line width=0.6mm, -{Latex[length=2.5mm, width=2.5mm]}] (-0.2,4.2) -- node[above=1mm] {${U}$} (-.3,4.3);
            \draw[black, line width=0.6mm, -{Latex[length=2.5mm, width=2.5mm]}] (4.2,4.2) -- node[above=1mm] {${V}$} (4.3,4.3);
            \draw[black, line width=0.2mm, <->, >=Latex] (0,0.5) -- 
            node[left=0.6mm] {$2\Delta \tau $} 
            (0,2);

            \node[green] at (2.3, 3.7) {\textbf{ C}};
            \fill[green] (2.57,3.43) circle (0.05);
            \node[green] at (1.5, 2.94) {\textbf{ B}};
            \node[magenta] at (2.08, 2.5) {\textbf{A}};

            \fill[green] (1.67,2.53) circle (0.05);
            \fill[magenta] (1.85,2.35) circle (0.05);

            \node[red] at (4.4, 0.3) {-$t_w$};
        \end{tikzpicture}
        \caption{SAdS$_D$ geometry (with $D>3$) deformed by a shock wave. }
        \label{fig:sads}
    \end{subfigure}
    \caption{Comparison of critical times for null geodesics in BTZ and SAdS$_D$ geometries deformed by a shock wave. The dashed lines represent null rays in the absence of a shock wave.
    Here, we are examining geodesics that start from the boundary at time $\tau_R=0$.
}
    \label{fig:comparison_combined}
\end{figure}

Instead of taking the infinite mass limit, we will now consider a generic mass $M$, SAdS black hole geometry.
In Kruskal coordinates the geometry is given by
\begin{align}
ds^2&=-\frac{ f(r)}{\kappa^2}e^{-2\kappa r_*(r)}dUdV + r^2 d\Omega_{D-2}^2 \, ,
\end{align}
with $r^*$ the usual tortoise coordinate, $\kappa$ the surface gravity due to a horizon at $r=r_b$. Let us now introduce a spherically symmetric shock wave perturbation of asymptotic (AdS boundary) energy $E \ll M$, emitted at a time $-t_w$ from the right asymptotic AdS boundary, implying that we take the radius of the natural observer to $r=r_\mathcal{O} \rightarrow \infty$. We define coordinates $\tilde{U},\tilde{V}$ to the right of the perturbation, and continue to use $U,V$ to the left. 
The shock wave shell propagates on the surface defined by
\begin{eqnarray}
\tilde{V}^w = e^{{\tilde{\kappa}}(\tilde{r}_*(r_\mathcal{O}) - t_w)} \, , \hspace{20pt} V^w = e^{\kappa(r_*(r_\mathcal{O}) - t_w)}\,,\label{u}    
\end{eqnarray}
The mass of the AdS black hole geometry jumps from $M$ to $M + E$ as one crosses the shock wave from the left to the right. Effectively we are gluing together two different SAdS geometries along a radial null surface. As a consequence, the points A and B in figure \ref{fig:sads} have the same radius $r = r_s$, but the null coordinate $U$ is shifted as one crosses the shock wave.

It is well known that introducing a shock wave shell close to the horizon of a three-dimensional BTZ black hole creates a characteristic time-delay when 
probing null geodesics cross the shock wave, see figure \ref{fig:btz-shock}. We will quantify this time-delay by computing the shift in the critical time at the left AdS boundary. 
First observe that without a shock wave, for $D>3$ AdS black holes, the Penrose diagram fails to be a perfect square due to the bending of the future (and past) singularity. As a result, in $D>3$ the critical time starts out non-vanishing and becomes larger after introducing the shock wave-deformed SAdS$_D$ geometry in $D>3$ dimensions.

We will now explicitly compute the  critical time before and after introducing shock waves. 
From the endpoints of the null geodesics, we can read off the corresponding Kruskal coordinates
\begin{align}
U&=-\exp\left(-\kappa(t_L-r^*(r_\mathcal{O}))\right),\label{eq-U-tL-method2}\\
\tilde V&=\exp\left(\tilde\kappa(t_R+\tilde r^*(r_\mathcal{O}))\right),\label{eq-V-tR-method2}
\end{align}
where we denote with $\tilde{\kappa}$ and $\tilde{r}^*(r)$ the fact that they correspond to a SAdS geometry with mass $M+E$ (because we are in the region 
$\tilde V > \tilde V^w$
) instead of $M$ (which corresponds to the region $V < V^w$). Note that $t_L=\tau_L-i\beta/2$ has acquired a constant imaginary part (See figure \ref{complextads}), and we denote the real part of this complex time as $\tau_L$. 
Since the shock wave starts at time $-t_w < 0$ we have 
\begin{equation}\label{eq-Vw-method2}
\begin{aligned}
V^w&=\exp(\kappa (-t_w+r^*(r_\mathcal{O})))\,,\\ 
\tilde V^w&=\exp(\tilde\kappa (-t_w+\tilde r^*(r_\mathcal{O})))\,.
\end{aligned}
\end{equation}
Next, consider the product $UV$ in the three different points $A$, $B$, $C$:
\begin{equation}\label{eq-points-ABC}
\begin{aligned}
UV^w&=-\exp\left(2\kappa r^*(r_s)\right)\,,\\
\tilde U\tilde V^w&=-\exp\left(2\tilde \kappa \tilde r^*(r_s)\right),\\
\tilde U\tilde V&=-\exp\left(2\tilde \kappa \tilde r^*(r_f)\right).
\end{aligned}
\end{equation}
In these expressions $r_f$ corresponds to the intersection of the null geodesic with the (future) singularity, implying that $r_f = 0$.
We introduced $r_s$ as the radius where the shock wave intersects at point A with the null geodesic emitted at time $\tau_L$ from the left AdS boundary at some fixed value of $U$. Points A and B share the same radius $r = r_s$, but the value of $U$ is shifted due to the shock wave. Starting from 
\begin{equation}
U \tilde V=\tilde U \tilde V \frac{U V^w}{\tilde U \tilde V^w} \frac{\tilde V^w}{V^w}\,,
\end{equation}
this implies, after using \eqref{eq-U-tL-method2} and \eqref{eq-V-tR-method2} to rewrite the left-hand-side, and equations \eqref{eq-Vw-method2} and \eqref{eq-points-ABC} to rewrite the right-hand-side
\begin{align}\label{eq-tR-tL}
\tilde\kappa t_R-\kappa t_L&
=2 \kappa ( r^*(r_f)- r^*(r_\mathcal{O}))-2\tilde \kappa \tilde r^*(r_s)+2\kappa r^*(r_s)\nonumber\\
&\quad\quad +2\tilde\kappa\tilde r^*(r_f)-2\kappa r^*(r_f) +t_w (\kappa-\tilde\kappa)\,.
\end{align}

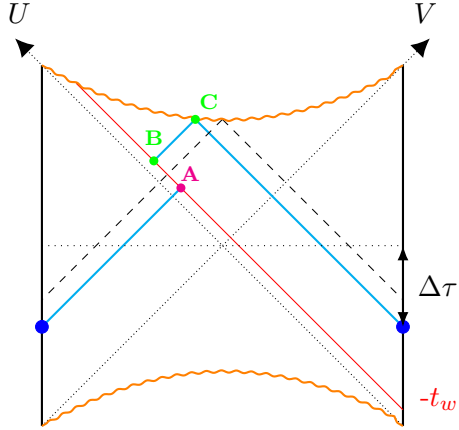
\begin{figure}[]
\centering
\begin{tikzpicture}[scale=1.2]
\draw[thick] (0,0) -- (0,4);
\draw[thick] (4,0) -- (4,4);

\draw[densely dotted] (0,0) -- (4.2,4.2);
\draw[densely dotted] (-0.2,4.2) -- (4,0);

\draw[thick, cyan] (1.7,3.4) -- (4,1.1);
\draw[thick, cyan] (1.54,2.64) -- (0,1.1);
\draw[thick, cyan] (1.24,2.94)-- (1.7,3.4);


\draw[black, line width=0.6mm, -{Latex[length=2.5mm, width=2.5mm]}] (4.2,4.2) -- node[above=1mm] {${V}$} (4.3,4.3);
\draw[red] (4,0.18)--(0.38,3.8);
\draw[black, line width=0.6mm, -{Latex[length=2.5mm, width=2.5mm]}] (-0.2,4.2) -- node[above=1mm] {${U}$} (-.3,4.3);

\draw[dashed, black] (2,3.4) -- (0,1.4);
\draw[dashed, black] (2,3.4) -- (4,1.4);

\draw[dotted] (0,2) -- (4,2);

\node[red] at (4.4, 0.3) {-$t_w$};

\filldraw[blue] (4,1.1) circle (2pt); 
\filldraw[blue] (0,1.1) circle (2pt) ;
\draw[black, line width=0.2mm, <->, >=Latex] (4,2) -- node[right=0.6mm] {$\Delta \tau$} (4,1.1);

\draw[thick, color=orange, decorate, decoration={snake,amplitude=.15mm,segment length=2mm, post length=.6mm,pre length=.6mm}] (0,4) to[out=-30, in=-150] (4,4);
\draw[thick, color=orange, decorate, decoration={snake,amplitude=.15mm,segment length=2mm, post length=.6mm,pre length=.6mm}] (0,0) to[out=30, in=150] (4,0);

\node[green] at (1.8, 3.6) {\scriptsize\textbf{ C}};
            \fill[green] (1.7,3.4) circle (0.05);

 \node[magenta] at (1.6, 2.8) {\scriptsize{ \textbf{A}}};
            \fill[magenta] (1.54,2.64) circle (0.05);

 \node[green] at (1.2, 3.2) {\scriptsize{ \textbf{B}}};
 \fill[green] (1.24,2.94) circle (0.05);

\end{tikzpicture}
\caption{
Symmetric geodesics (i.e. those with $\tau_R=-\tau_L$) in a SAdS black hole deformed by a shock wave. The dashed lines represent the symmetric geodesic in the absence of a shock wave.
 }
\label{fig:SAdspenrose-shock wave-symmetric}
\end{figure}

We will now introduce the double scaling limit, which implies that for a fixed mass $M$ of the black hole, we will assume
\beq \label{eq-doublescalinglimit}
\frac{E}{M}\to 0 \quad\text{and}\quad t_w\to +\infty\quad\text{such that}\quad \frac{E}{M} e^{\kappa t_w }\quad\text{is fixed}\, .
\eeq 
In this limit, to leading order in $E/M$ , we find,  
\beq 
\frac{\tilde V^w}{V^w}\simeq 1\,.
\eeq 
As a consequence, again to leading order, the shift in $U$ when we cross the surface $V=V^w$ equals {\cite{Shenker:2013pqa}}
\begin{align}\label{eq-alpha-prev}
\alpha&=\tilde{U}-U\simeq \frac{\tilde U \tilde V^w-U V^w}{V^w}=-\frac{1}{V^w}\left(\exp\left(2\tilde\kappa \hs \tilde r^*(r_s)\right)-\exp\left(2\kappa \hs  r^*(r_s)\right)\right)\nonumber\\
&\simeq -\frac{1}{V^w} E \frac{d}{dM}\left(\exp\left(2\kappa \hs  r^*(r_s)\right)\right).
\end{align}
Notice that we can write the exponential in the last line as
\beq \label{eq-def-C}
 \exp\left(2\kappa r^*(r_s)\right)=-U V^w=  (r_s-r_b) C(r_s,r_b) \,,
\eeq 
where $C(r_s, r_b)$ is a smooth and regular function that is positive for 
any $r_s> 0$.
For the sake of illustration, consider the case of a $D=5$ dimensional SAdS spacetime with a generic mass $M$, where the blackening factor, with the AdS radius normalized to $\ell=1$, is given by  
\[
f(r)=1+r^2-\frac{r_b^2(1+r_b^2)}{r^2}\,.
\]  
It is straightforward to derive that the function $C(r_s, r_b)$ for this case is
\[
C(r_s, r_b)\Big|_{D=5}=\frac{1}{r_s+r_b}\exp\left(2\sqrt{1+\frac{1}{r_b^2}}\hs\arctan\left(\frac{r_s}{\sqrt{1+r_b^2}}\right)\right).
\]  
Thus, we can write the shift in \eqref{eq-alpha-prev} in terms of this smooth function $C$ as
\begin{align}\label{eq-alpha}
\alpha&=\tilde{U}-U\simeq -\frac{1}{V^w} E \frac{d}{dM}\left((r_s-r_b(M))C(r_s,r_b(M))\right)\nonumber\\
&\simeq \frac{E}{V^w}  \hs \frac{dr_b}{dM}\hs C(r_b,r_b)\,,
\end{align}
where in the last line we neglected terms of order $\mathcal{O}(r_s-r_b)$, which vanish in the limit $t_w\to \infty$. As $E/V^w\propto E e^{\kappa t_w}$ we find that the shift $\alpha$ at $r\sim r_b$ is constant in the double scaling limit \eqref{eq-doublescalinglimit}. Following \cite{Shenker:2013pqa}, we can use the first law of thermodynamics to write the derivative $dr_b/dM$ in terms of the temperature $T=dM/dS_{BH}$ and the (derivative of the) entropy. This allows us to express the shift in the Kruskal coordinate $U$ as
\be
\alpha\simeq \frac{E}{V^w}  \hs C(r_b,r_b)\hs \frac{4G_N}{T A'(r_b)}=\frac{E}{V^w}\hs C(r_b,r_b)  \hs \frac{ r_b}{(D-2)\hs T \hs S_{BH}}\,,
\ee 
where $A(r_b)=\Omega_{D-2} r_b^{D-2}$ is the area of the black hole horizon and $S_{BH}=\displaystyle\frac{A(r_b)}{4G_N}$ is its entropy. 

Notice that in the double-scaling limit we can write the last two terms of the first line of \eqref{eq-tR-tL} as
\begin{align}
2\tilde\kappa \tilde r^*(r_s)-2\kappa r^*(r_s)
&=\log\left(\frac{\tilde U \tilde V^w}{U V^w}\right)\simeq  \log\left(\frac{\tilde U}{U}\right)\nonumber\\
& \simeq \log\left(1+\frac{\alpha}{U}\right)  , 
\end{align}
where we used \eqref{eq-def-C} and that $\tilde\kappa\simeq \kappa$, $\tilde U=U+\alpha$ to leading order. Furthermore, the second line of \eqref{eq-tR-tL} vanishes in this limit, and as a result the time difference \eqref{eq-tR-tL} equals
\begin{equation}
t_R-t_L\simeq 2(r^*(r_f)-r^*(r_\mathcal{O}))-\frac{1}{\kappa }\log\left(1+\frac{\alpha}{U}\right),
\end{equation}
which implies that the critical time associated with the black hole horizon in AdS is 
\begin{eqnarray}\label{eq-criticaltime-U}
\Delta \tau &\equiv&\frac{\tau_R-\tau_L}{2} 
\nonumber\\
&=&\text{Re}\left(r^*(r_f)-r^*(r_\mathcal{O})\right)-\frac{1}{2\kappa }\log\left(1+\frac{\alpha}{U}\right)\nonumber\\
&=&-\mathcal{T} -\frac{1}{2\kappa }\log\left(1+\alpha\hs e^{\kappa (\tau_L-r^*(r_{\mathcal{O}}))}\right) .
\end{eqnarray}
For an AdS black hole, $r_f\rightarrow0$ while $r_\mathcal{O}\rightarrow\infty$ 
\cite{Festuccia:2005pi}. The first term corresponds to the contribution to the critical time without a shock wave, implying a Penrose diagram that is not a perfect square: the past and future singularities are bent inwards, with respect to the two straight vertical lines that represent the AdS boundaries. The second term, of order $\alpha$, is due to the insertion of a shock wave and depends on $U$, which in turn is fixed by the emission time of the null geodesic (see equation \eqref{eq-U-tL-method2}). Both contributions are strictly negative quantities for any $D>3$ {provided $\alpha>0$ (i.e. that the shock waves satisfy NEC)}. In the limit of infinitely massive AdS black holes, without a shock wave, the expression reduces to equation \eqref{T}.  

For a symmetric null geodesic $\tau_R=-\tau_L= \Delta \tau$ (see figure \ref{fig:SAdspenrose-shock wave-symmetric}), so we can replace $\tau_L=- \Delta \tau$ in equation \eqref{eq-criticaltime-U}, and the critical time is given by the solution to the equation 
\be 
\Delta \tau=-\mathcal{T} -\frac{1}{2\kappa }\log\left(1+\alpha\hs e^{-\kappa (\Delta \tau+r^*(r_{\mathcal{O}}))}\right)  .\label{adsbhsymm-eq}
\ee
This can easily be translated to a quadratic equation for the variable $ \exp(-\kappa\Delta \tau)$, which in turn yields only one solution with $\Delta \tau$ real that corresponds to the following critical time
\be 
\Delta \tau=-\mathcal{T} -\frac{1}{\kappa }\log\left(\sqrt{1+\frac{\alpha^2}{4}}+\frac{\alpha}{2}\right)  ,\label{adsbhsymm}
\ee 
where we used that $\text{Re}(r^*(r_f))=0$, given our definition of the tortoise coordinate in \eqref{equ:rstar-SAdS}.
For a BTZ black hole ($D=3$) the first term vanishes and in that case \eqref{adsbhsymm} coincides with equation (3.22) of \cite{Anegawa:2023dad} in the small $\alpha$ limit, where they also considered symmetric geodesics. The time difference introduced by the shock wave is proportional to the inverse temperature $\beta=2\pi/\kappa$ of the black hole. Because the set-up is time-symmetric, the critical time for null geodesics bouncing off the past singularity just carries a different sign\footnote{To be precise, applying a rotation $t\mapsto t-i\beta/2$ sends the symmetric null geodesic bouncing off the future singularity to the symmetric null geodesic bouncing off the past singularity. Note that this transformation preserves the shock wave geometry in the double scaling limit, and that it simply changes the sign of the critical time $(\tau_R-\tau_L)/2$ because it flips $\tau_R\leftrightarrow \tau_L$.}, so the onset of linear growth in the complexity parameter is delayed (for positive times), which is known as the switchback effect. The duration of the phase of relatively slow evolution of the complexity parameter is therefore extended due to the introduction of the shock wave perturbation, and equal to twice the computed critical time. The precise magnitude crucially depends on the value of $\alpha$ in the double scaling limit. Note that this time delay is expected to be related to the scrambling time,
which is obtained by evaluating the time $t_w$ at the right AdS boundary at which a minimal energy perturbation $E\sim 1/\beta$ gives rise to an order one shift $\alpha\sim\mathcal{O}(1)$ close to the horizon. Looking at the double scaling limit (recall equation \eqref{eq-doublescalinglimit}) this gives $t_w \propto \beta \log (\beta M)=\beta \log S$. The result for the switchback delay is also proportional to the inverse temperature. The logarithmic factor is of order $\alpha$ in the limit of small shock wave perturbations, and of order one for (large) order one shifts, so much shorter than a scrambling time.

\subsection{Results for perturbed critical times in a Schwarzschild-de Sitter geometry}

\label{krisko}
Having briefly reviewed the shock wave switchback delay in the AdS context, we are ready to discuss the SdS geometry \cite{Anegawa:2023dad, Baiguera:2023tpt, Baiguera:2024xju, Aguilar-Gutierrez:2023pnn}. As we emphasized already, throughout this work we will assume that we can split the SdS geometry in an interior black hole region and an exterior de Sitter region, glued together at the static sphere where we assume reflecting Dirichlet boundary conditions, resulting in a semi-classical gravitational state that is (approximately) stable. This a priori might suggest we can compute the critical time differences due to shock waves (satisfying the Null Energy Condition (NEC)) for the interior black hole and exterior de Sitter region independently. However, as already alluded to, in the SdS geometry it is not consistent, due to its spatial compactness, to introduce a single positive energy radial shock wave. Instead, consistent gluing of the SdS geometries with slightly different masses implies that when emitting a positive energy shock wave through the cosmological horizon, one needs to also send a negative energy shock wave into the black hole \cite{Aalsma:2021kle} (see figure \ref{fig:SdS-setup}). This conserves local energy-momentum, which in the absence of an asymptotic boundary and due to the compact nature of the spatial sections is the only consistent way to introduce a perturbation in the SdS geometry. Doing so implies we effectively couple the interior black hole and exterior de Sitter region. Keeping this in mind, let us first just consider the effect of introducing a positive energy shock wave in the exterior de Sitter region. 

\subsubsection{The exterior asymptotic dS region}

\begin{figure}[h!]
    \centering
    \begin{subfigure}[t]{0.45\textwidth}
        \centering
        \begin{tikzpicture}[scale=1]
            \draw[dashed, blue] (4,2) -- (2,4);
            \draw[dashed, blue] (1.05,3.05) -- (2,4);
            \draw[thick] (4,0) -- (4,4);
            \draw[thick] (0,0) -- (0,4);
            \draw[densely dotted] (4,0) -- (-0.2,4.2);
            \draw[violet] (4,4) -- (0,4);
            \draw[densely dotted] (4.2,4.2) -- (0,0);
            \draw[red] (4,0.1) -- (0.1,4);
            \draw[thick, blue] (0,2) -- (1.05,3.05);
            \draw[thick, blue] (1.3,2.8) -- (2.5,4);
            \draw[thick, blue] (4,2.5) -- (2.5,4);
            \draw[dotted] (4,2) -- (0,2) node[pos=1.02,left]{$\tau_L=0$};
            \draw[black, line width=0.6mm, -{Latex[length=2.5mm, width=2.5mm]}] (4.2,4.2) -- node[above=1mm] {${U}_c$} (4.3,4.3);
            \draw[black, line width=0.6mm, -{Latex[length=2.5mm, width=2.5mm]}] (-0.2,4.2) -- node[above=1mm] {${V}_c$} (-.3,4.3);
            \node[green] at (2.7, 4.3) {\textbf{C}};
            \node[green] at (1.8, 2.8) {\textbf{B}};
            \node[magenta] at (0.7, 3.0) {\textbf{A}};
            \fill[green] (2.5,4) circle (0.05);
            \fill[magenta] (1.05,3.05) circle (0.05);
            \fill[green] (1.3,2.8) circle (0.05);
            \node[red] at (4.35, 0.1) {$-t_w$};
            \draw[black, line width=0.2mm, <->, >=Latex] (4,2.5) -- node[right=0.6mm] {$2\Delta \tau_c$} (4,2);
            \draw[violet] (4,0) -- (0,0);
        \end{tikzpicture}
        \caption{Penrose diagram of empty dS spacetime showing the time advance $\Delta \tau_c$ for light particles in the presence of a shock wave.}
        \label{fig:dstimeadv}
    \end{subfigure}
    \hfill
    \begin{subfigure}[t]{0.45\textwidth}
        \centering
        \begin{tikzpicture}[scale=0.8]
            \draw[black, line width=0.6mm, -{Latex[length=2.5mm, width=2.5mm]}] (4.7,4.7) -- (4.9,4.9);
            \draw[black, line width=0.6mm, -{Latex[length=2.5mm, width=2.5mm]}] (-.1,4.6) -- (-.3,4.8);
            \draw[densely dotted] (0,0)--(4.7,4.7);
            \draw[densely dotted] (4.5,0)--(0-.1,4.6);
            \draw[red] (4.5,0.17) -- (0.1,4.57);
            \node[above] at (4.5,-0.7) {\scriptsize $r=r_{\mathcal{O}}$};
            \draw[thick] (0,0) -- (0,4.5);
            \draw[thick] (4.5,0) -- (4.5,4.5);
            \node[above] at (0,-0.7) {\scriptsize $r=r_{\mathcal{O}}$};
           
            \node[above] at (2.2,5.2) {$\mathcal{I}^+$};
            \node[below] at (2.2,-0.6) {$\mathcal{I}^-$};
            \node[above] at (-0.4,4.7) {$V_c$};
            \node[above] at (5.2,4.5) {$U_c$};
            \draw[thick,blue] (0,2.25) -- (1.21,3.46);
            \draw[thick,blue] (3.32,4.99) -- (1.5,3.17);
            \draw[thick,blue] (3.32,4.99) -- (4.5,3.81);
            \draw[blue,dashed] (2.85,5.1)--(1.21,3.46);
            \draw[blue,dashed] (2.85,5.1)--(4.5,3.45);
            \draw[black, line width=0.2mm, <->, >=Latex] (4.5,3.81) -- node[right=0.6mm] {$2\Delta \tau_c$} (4.5,2.25);
            \draw[thick, color=violet] (4.5,4.5) to[out=151, in=29] (0,4.5);
            \draw[thick, color=violet] (4.5,0) to[out=-151, in=-29] (0,0);
            \node[green] at (3.43, 5.3) {\textbf{ C}};
            \node[green] at (2.0, 3.1) {\textbf{ B}};
            \node[magenta] at (1.1, 3.9) {\textbf{ A}};
            \fill[green] (3.32,4.99) circle (0.05);
            \fill[magenta] (1.21,3.46) circle (0.05);
            \fill[green] (1.5,3.17) circle (0.05);

            \draw[dotted] (4.5,2.25) -- (0,2.25) node[pos=1.02,left]{$\tau_L=0$};
            \node[red] at (5, 0.2) {$-t_w$};
        \end{tikzpicture}
        \caption{Shock wave through SdS spacetime from the static sphere towards the asymptotic cosmological region.}
        \label{fig:shock wave}
    \end{subfigure}
    \caption{ Penrose diagram of asymptotically dS spacetimes with a shock wave along $U_c=U_c^w$. These geometries are: (a) empty dS spacetime (b) asymptotic dS region of SdS.
    }
    \label{fig:merged_diagrams}
\end{figure}
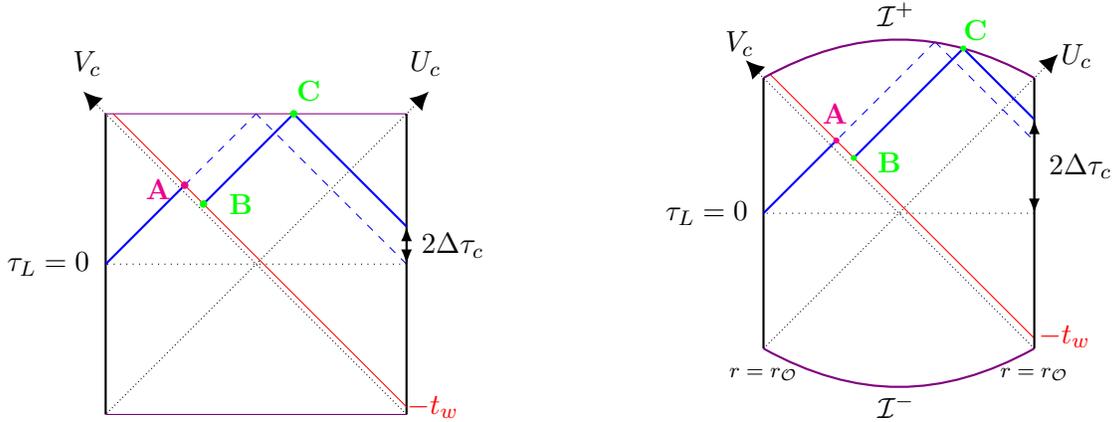

 As before, we will introduce Kruskal coordinates to understand the shock wave solutions in SdS spacetime with respect to an observer located on the static sphere. We will emit a positive energy shock wave from the static sphere through the cosmological horizon, towards future spacelike infinity. This means we have to glue together an SdS geometry with mass $M+E$ to one with mass $M$, along some null direction. We can write the metric in the asymptotic dS region of SdS in terms of the Kruskal coordinates that we introduced in Section~\eqref{tor1}
\begin{eqnarray}
    ds^2 = -\frac{(r-r_b)^{\frac{\kappa_c}{\kappa_b}+1}(r+r_b+r_c)^{-\frac{\kappa_c}{\kappa_b}+2}}{\kappa_c^2 \ell^2 r} \, dU_c dV_c 
    + r^2 d\Omega_{d-1}^2\,.
    \label{eq-sds-metric}
\end{eqnarray}
Let us now introduce a spherically symmetric shock wave perturbation emitted from the right static sphere at time \( t = -t_w \), corresponding to a shell propagating along the surface $U_c=U_c^w$. 
This means that the SdS mass equals $M$ for $U_c>U_c^w$ (i.e. after the shock wave has been emitted on the right static sphere) and $M+E$ in the region to the left of the radial positive shock wave with (static sphere) energy $E$, which includes the left static sphere. 
A null geodesic intersecting the shock will experience a Shapiro time advance, which in an asymptotic de Sitter geometry implies that some null geodesics can now reach the conjugate static sphere. For more details, see Refs.~\cite{Aalsma:2020aib,Anegawa:2023dad,Geng:2020kxh}. In figures~\ref{fig:dstimeadv} and \ref{fig:shock wave} we illustrated how shock waves cause a time advance in a pure and Schwarzschild-de Sitter geometry. 

For pure de Sitter, due to the perfect square nature of the Penrose diagram, in the absence of shock waves the critical time vanishes for a symmetric null geodesic connecting the two poles, bouncing off future spacelike infinity. However, after introducing a shock wave with positive energy $E$, the critical time is now non-vanishing \cite{Anegawa:2023dad}, as should be clear from figure~\ref{fig:dstimeadv}, which can be traced back to the fact that introducing positive energy density in de Sitter space elongates the Penrose diagram. We will be interested in quantifying the time difference of this critical time in the exterior de Sitter region of the SdS geometry, connected to the outward bending of de Sitter spacelike infinity, after the introduction of a positive energy radial shock wave \cite{Anegawa:2023dad}. We will study the emission of an (outward) radial shock wave from the right static sphere radius $r=r_\mathcal{O}$. 

From the intersection of the shock wave with the right static sphere, we can read off the corresponding Kruskal coordinates 
\begin{align}
U_c&=\exp\left(\kappa_c(t_R-r^*(r_\mathcal{O}))\right),\label{eq-U-tL-method2-sds}\\
\tilde V_c&=-\exp\left(-\tilde\kappa_c(t_L+\tilde r^*(r_\mathcal{O}))\right),\label{eq-V-tR-method2-sds-x}
\end{align}
where we denote $\tilde{\kappa}_c$ and $\tilde{r}^*(r)$ with a tilde when we are in the region $\tilde U_c<\tilde U_c^w$ defining the SdS geometry with mass $M+E$, while we denote $\kappa_c$ and $r^*(r)$ without a tilde when we are in the SdS region $U_c>U_c^w$ with mass $M$.
 Note that $t_L$ has a (constant) imaginary part (See figure \ref{ccc}) and we define the real part as $\tau_L=\text{Re}(t_L)$ (and we use the same notation for $\tau_R=\text{Re}(t_R)=t_R$).
Recall that, since the shock wave starts at a time $t=-t_w$, where $t_w> 0$:
\begin{equation}\label{eq-Vw-method2-sds}
\begin{aligned}
{U}^w_c =e^{-{\kappa}_c( t_w+r_*( r_\mathcal{O}))}\,,\\    
\tilde{U}^w_c =e^{-\tilde {\kappa}_c( t_w+\tilde r_*(r_\mathcal{O}))}\,.
\end{aligned}
\end{equation}
Considering the points $A$, $B$ and $C$, we derive  
\begin{equation}\label{eq-points-ABC-sds}
\begin{aligned}
\tilde U_c^w\tilde V_c&=-\exp\left(-2\tilde\kappa_c \tilde r^*(r_s)\right),\\
 U_c^w V_c&=-\exp\left(-2 \kappa_c  r^*(r_s)\right),\\
U_c V_c&=- \exp\left(-2 \kappa_c  r^*(r_f)\right).
\end{aligned}
\end{equation}
Here $r_f$ is the endpoint of the null geodesic starting from the right static sphere $r=r_\mathcal{O}$ and as we are considering outward (exterior) null geodesics they will reach future spacelike infinity, hence $r_f\rightarrow\infty$. We further define $r_s$ as the radius where the shock wave intersects with the null geodesic at fixed $\tilde V_c$, emitted from the left static sphere (point A). Points A and B share the same radius $r = r_s$, but the value of $V_c$ is shifted due to the shock wave. Starting from
\begin{equation}
U_c \tilde V_c=U_c  V_c \frac{\tilde U_c^w \tilde V_c}{ U_c^w  V_c} \frac{ U_c^w}{\tilde U_c^w}\,,
\end{equation}
implies, after using \eqref{eq-U-tL-method2-sds} and \eqref{eq-V-tR-method2-sds-x} to rewrite the left-hand-side, and equations \eqref{eq-Vw-method2-sds} and \eqref{eq-points-ABC-sds} to rewrite the right-hand-side, 
\begin{align}\label{eq-tR-tL-sds}
\kappa_c t_R-\tilde\kappa_c t_L&
=2\kappa_c ( r^*(r_{\mathcal{O}})- r^*(r_f))-2\tilde \kappa_c \tilde r^*(r_s)+2\kappa_c r^*(r_s)\nonumber\\
&\quad\quad +2\tilde\kappa_c\tilde r^*(r_{\mathcal{O}})-2\kappa_c r^*(r_{\mathcal{O}}) +t_w (\tilde\kappa_c-\kappa_c)\,.
\end{align}
After taking the double scaling limit
\beq \label{eq-doublescalinglimit-sds}
\frac{E}{M}\to 0 \quad\text{and}\quad t_w\to +\infty\quad\text{such that}\quad \frac{E}{M} e^{\kappa_c t_w}\quad\text{is fixed}\,,
\eeq 
this implies at leading order 
\beq 
\frac{\tilde U^w_c}{U^w_c}\simeq 1\,.
\eeq 
Moreover, it is easy to see that the shift in $V_c$ when we cross the surface $U_c=U_c^w$ is constant in the double scaling limit:
\begin{align}\label{eq-alpha-sds-prev}
\alpha_c&=\tilde{V}_c-V_c\simeq \frac{(\tilde U_c^w \tilde V_c-U_c^w V_c)}{U_c^w}=-\frac{1}{U_c^w}\left(\exp\left(-2\tilde\kappa_c \hs \tilde r^*(r_s)\right)-\exp\left(-2\kappa_c \hs  r^*(r_s)\right)\right)\nonumber\\
&\simeq -\frac{1}{U_c^w} E 
\frac{d}{dM}\left(\exp\left(-2\kappa_c \hs  r^*(r_s)\right)\right).
\end{align}
In analogy to the asymptotic AdS calculation, we can then write the exponential in the last line as
\beq \label{eq-def-C-sds}
\exp\left(-2\kappa_c r^*(r_s)\right)= -U_c V_c^w=(r_c-r_s) C_c(r_s,r_c)\,,
\eeq 
where $C_c(r_s, r_c)$ is a smooth and regular function that is positive for 
any $r_s> r_b$.
For the sake of illustration, we can easily see from equation \eqref{eq-toget-Cc} that when $D=4$ this function is given by
\be
C_c(r_s, r_c)\Big|_{D=4}=\frac{1}{r_s+r_b+r_c}\left(\frac{r_s+r_b+r_c}{r_s - r_b}\right)^{\kappa_c/\kappa_b}.
\ee
Thus, we can write \eqref{eq-alpha-sds-prev} in terms of $C_c$ as
\begin{align}\label{eq-alpha-sds}
\alpha_c&=\tilde{V}_c-V_c\simeq -\frac{E}{U_c^w} \frac{d}{dM}\left((r_c( M)-r_s)C_c(r_s,r_c(M))\right)\nonumber\\
&\simeq \frac{E}{U_c^w}  \hs \left|\frac{dr_c}{dM}\right|\hs C_c(r_c,r_c)\,,
\end{align}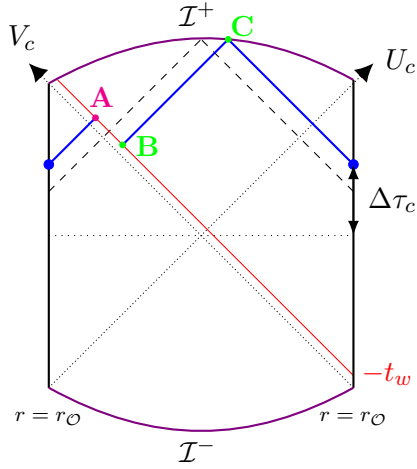
\begin{figure}[]
    \centering
        \begin{tikzpicture}[scale=0.9]

        \draw[dashed, black]  (0,2.9) -- (2.25,5.15);
            \draw[dashed, black]  (4.5,2.9) -- (2.25,5.15);
            
              \draw[black, line width=0.6mm, -{Latex[length=2.5mm, width=2.5mm]}] (-0.1,4.6) -- (-.3,4.8);
            \draw[black, line width=0.6mm, -{Latex[length=2.5mm, width=2.5mm]}] (4.6,4.6) -- (4.8,4.8);
            \draw[dotted] (0,2.25) -- (4.5,2.25);
            \draw[densely dotted] (0,0) -- (4.6,4.6);
            \draw[densely dotted] (4.5,0) -- (-.1,4.6);
            \draw[red] (4.5,0.18) -- (0.13,4.55);
            \draw[thick] (0,0) -- (0,4.5);
            \draw[thick] (4.5,0) -- (4.5,4.55);
            \draw[thick, color=violet] (4.5,4.55) to[out=151, in=29] (0,4.5);
            \draw[thick, color=violet] (4.5,0) to[out=-151, in=-29] (0,0);
            \node[above] at (2.2,5.2) {$\mathcal{I}^+$};
            \node[below] at (2.2,-0.6) {$\mathcal{I}^-$};
           \node[above] at (-0.4,4.85) {$V_c$};
    \node[above] at (5.2,4.5) {$U_c$};
            \node[above] at (4.5,-0.7) {\scriptsize $r=r_{\mathcal{O}}$}; 
            \node[above] at (0,-0.7) {\scriptsize $r=r_{\mathcal{O}}$}; 
            \node[red] at (5, 0.2) {$-t_w$};
            \draw[thick,blue] (0,3.3) -- (0.69,3.99);
            \draw[thick,blue] (1.09,3.59) -- (2.65,5.15);
            \draw[thick,blue] (4.5,3.3) -- (2.65,5.15);
            \node[green] at (2.78, 5.35) {\textbf{ C}};
            \node[green] at (1.38, 3.57) {\textbf{ B}};
            \node[magenta] at (0.7, 4.3) {\textbf{ A}};
            \fill[green] (1.09,3.59) circle (0.05);
            \fill[green] (2.65,5.15) circle (0.05);
            \fill[magenta] (0.69,3.99) circle (0.05);
\filldraw[blue] (4.5,3.3) circle (2pt); 
\filldraw[blue] (0,3.3) circle (2pt); 

\draw[black, line width=0.2mm, <->, >=Latex] (4.5,2.25) -- node[right=0.6mm] {$\Delta \tau_c$} (4.5,3.3);

        \end{tikzpicture}
    \caption{Symmetric geodesics (i.e. those with $\tau_R=-\tau_L$) in the exterior SdS 
        region deformed by a shock wave. The dashed lines represent the symmetric geodesic in the absence of a shock wave. 
}
    \label{fig:shock_wavessdsxxx}
\end{figure}
In the final step, we ignored terms of order \(\mathcal{O}(r_c - r_s)\) that vanish in the limit \(t_w \to \infty\). Additionally, we utilized the fact that \( \frac{dr_c}{dM} < 0 \).
 The final result is therefore proportional to the fixed combination $E/U_c^w\propto E e^{\kappa_c t_w}$ in the double scaling limit \eqref{eq-doublescalinglimit-sds}, hence it is a constant. In analogy with our previous calculation for asymptotic AdS, we can use the first law of thermodynamics in the cosmological region (which has a minus sign) to write the derivative $dr_c/dM$ in terms of the temperature $T_c=-dM/dS_{c}$ and the (derivative of the) entropy. This allows us to express the shift in the Kruskal coordinate $V_c$ as
\be\label{eq-alphac-Sc}
\alpha_c\simeq \frac{E}{U_c^w}  \hs C_c(r_c,r_c)\hs \frac{4G_N}{T_c A'(r_c)}=\frac{E}{U_c^w}\hs C_c(r_c,r_c)  \hs \frac{ r_c}{(D-2)\hs T_c\hs S_{c}}\,,
\ee 
where $A(r_c)=\Omega_{D-2} r_c^{D-2}$ is the area of the cosmological horizon and $S_{c}=\displaystyle\frac{A(r_c)}{4G_N}$ is its entropy. 

Let us now apply the double scaling limit to \eqref{eq-tR-tL-sds}. In particular, the last two terms of the first line can be written  in this limit as
\begin{align}
-2\tilde\kappa \tilde r^*(r_s)+2\kappa r^*(r_s)
&= -\log\left(\frac{ U_c^w V_c}{\tilde U_c^w \tilde V_c}\right)\simeq  -\log\left(\frac{ V_c}{\tilde V_c}\right)\nonumber\\
&\simeq -\log\left(1-\frac{\alpha_c}{\tilde V_c}\right),
\end{align}
where we used \eqref{eq-def-C-sds}, $\tilde\kappa_c\simeq \kappa_c$ and $V_c=\tilde V_c-\alpha_c$ to leading order. Since the second line of equation \eqref{eq-tR-tL-sds} vanishes in the double scaling limit, this time difference therefore reads
\begin{equation}
t_R-t_L=2(r^*(r_\mathcal{O})-r^*(r_f))-\frac{1}{\kappa_c }
\log\left(1-\frac{\alpha_c}{\tilde V_c}\right)
,
\end{equation}
which implies that the critical time associated with the exterior asymptotic dS region is  
\begin{align}
\Delta \tau_c&\equiv \frac{\tau_R - \tau_L}{2} 
= \frac{t_R - t_L + i \beta_c / 2}{2} \nonumber \\
&\simeq  \text{Re}\big[r^*(r_\mathcal{O}) - r^*(r_f)\big] 
-\frac{1}{2\kappa_c}\log\left(1 - \frac{\alpha_c}{\tilde V_c}\right) \nonumber \\
&\simeq   \mathcal{T}_c 
-\frac{1}{2\kappa_c}\log\left(1 - \alpha_c\hs e^{ \kappa_c (\tau_L +  r^*(r_{\mathcal{O}}))}\right) .\label{rupahimu}
\end{align}
Here {$r_f\to\infty$} and we have used that the imaginary part of $r^*(r_f) - r^*(r_{\mathcal O})$ is precisely $\beta_c/4$ \cite{Festuccia:2005pi}. 
The additional term {with the logarithm} 
is due to the shock wave emitted from the right static sphere, which in essence is the same as for pure dS spacetime \cite{Anegawa:2023dad}. The first term depends on the particular value of the radius and the presence of a black hole, i.e. in the absence of a black hole
the first term will vanish. For any other radius and non-vanishing mass it is positive.  
 However, without proper normalization the critical time will diverge in the Nariai limit. For example, as is clear from \eqref{200}, $\kappa_c\rightarrow0$ in the Nariai limit, meaning that the term proportional to $\alpha_c$ in \eqref{rupahimu} blows up. Similarly, the first contribution 
 $\mathcal{T}_c$ also diverges in the Nariai limit, due to $r^*$, as seen from \eqref{tort}. Introducing the proper static sphere normalization \eqref{rupahimu} will remove these divergences, as we will now discuss in some detail.
 
For observers on the static sphere the proper time is measured by $t^{\mathcal{O}}=\gamma_{\mathcal{O}}\hs t$, as the Killing vector associated with such observers is identified with $\partial_{t^{\mathcal{O}}}$ \cite{morvan:2022rn,Faruk:2023uzs}.
As a consequence, the proper critical time measured on the static sphere is given by 
\begin{eqnarray}\Delta \tau_c^{\mathcal{O}} = \gamma_{\mathcal{O}}\hs\Delta \tau_c\,.
\end{eqnarray}
So we conclude that the critical time in the SdS exterior region, as measured by static sphere observers equals 
 \begin{eqnarray}\label{eq-taucO-tauL}
\Delta \tau_c^{\mathcal{O}}=\mathcal{T}_c^{\mathcal{O}}
-\frac{1}{2\kappa_c^{\mathcal{O}}}\log\left(1 - \alpha_c\hs e^{ \kappa_c (\tau_L +  r^*(r_{\mathcal{O}}))}\right) ,\end{eqnarray}
where we identified
\be 
\mathcal{T}_c^{\mathcal{O}}\equiv \gamma_{\mathcal{O}} \hs\mathcal{T}_c=\gamma_{\mathcal{O}}\hs\text{Re}\left(r^*(r_{\mathcal{O}}) - r^*(r_f)\right).
\ee 
The first term is strictly positive for any SdS mass, except in the Nariai and empty dS limit (where it vanishes), because $\lim_{r_f\rightarrow\infty}{\text{Re}(}r^*(r_f))=0$ and the real part of $r^*(r_\mathcal{O})$ is always a positive definite quantity for $r_b<r_\mathcal{O}<r_c$. 
Focusing on symmetric null configurations (see figure \ref{fig:shock_wavessdsxxx}), we can replace $\tau_L=- \Delta \tau_c$ in \eqref{eq-taucO-tauL}, resulting in an equation for the critical time $\Delta \tau_c$ (we will add the $\gamma_{\mathcal{O}}$ factor later)
 \begin{eqnarray}\label{eq-tauc-implicit}
\Delta \tau_c=\mathcal{T}_c-\frac{1}{2\kappa_c}\log\left(1 - \alpha_c\hs e^{ -\kappa_c (\Delta \tau_c-  r^*(r_{\mathcal{O}}))}\right).\end{eqnarray}
This can easily be translated to a quadratic equation for $e^{ -\kappa_c \Delta \tau_c} $, which in turn yields only one solution for $\Delta \tau_c\in\mathbb{R}$. The corresponding critical time measured by the static sphere observer (i.e. after restoring again the $\gamma_{\mathcal{O}}$ factor) is given by 
\be
 \Delta \tau_c^{\mathcal{O}}
 =\mathcal{T}_c^{\mathcal{O}}+\frac{1}{\kappa_c^{\mathcal{O}}}\log\left(\sqrt{1+\frac{\alpha_c^2}{4}}+\frac{\alpha_c}{2}\right).
 \label{sdsbhsymm-cosmo}
\ee
The first contribution, due to the nontrivial bending of dS spacelike infinity, distinguishes this from the pure dS geometry.  
We see that introducing a positive energy shock wave perturbation produces a characteristic delay in the critical time, as for the AdS black hole. However, in the asymptotic AdS case the delay is caused by a shrinking Wheeler-de Witt (WdW) wormhole volume, whereas here the WdW volume in between the two conjugate asymptotic dS static patches increases and in fact opens up the wormhole, see figure \ref{fig:shock_wavessdsxxx}, allowing information to be exchanged between two conjugate static sphere observers \cite{Aalsma:2020aib,Geng:2020kxh}. The detailed dependence on the shock wave energy of this so-called switchback delay is hidden in the logarithmic factor containing the shift $\alpha_c$, but it is clearly proportional to the static sphere inverse temperature $1/T_c^\mathcal{O}$ of the cosmological horizon and is considerably faster than the related scrambling time, being order one for order one shifts $\alpha_c$. The duration of the complexity plateau is extended by twice this delay. Interestingly, as also observed in \cite{Baiguera:2023tpt}, the origin of the linear growth in the exterior dS asymptotic region can be traced to the tip of the WdW causal diamond crossing spacelike future infinity, whereas for the AdS black hole the linear growth starts after the base (or bottom tip) of the WdW causal diamond detaches from the past singularity.

\subsubsection{The interior black hole region of SdS geometry}

As already alluded to, in the SdS geometry it is not consistent to add a single radial shock wave. Instead we can consistently introduce a pair of shock waves with a total energy that vanishes, see figure \ref{fig:SdS-setup}. Since we have studied positive energy shock waves passing through the exterior de Sitter region, we will now study the effect of emitting a negative energy shock wave into the interior black hole region. We can already anticipate that the resulting delay in the exterior de Sitter region should be accompanied by a comparable advance (earlier time, shorter duration) of the complexity plateau in the black hole region. This is consistent with a holographic dual picture where the maximum number of degrees of freedom of pure de Sitter are distributed in two separated (coarse-grained) thermodynamical reservoirs corresponding to the black hole and exterior de Sitter region. By introducing a pair of shock waves, with vanishing total energy, and negative energy emitted into the black hole, one is effectively reducing the mass of the black hole and moving degrees of freedom from the black hole to the exterior de Sitter region. Consequently, the macroscopic entropy and maximal limiting complexity related to the start of linear growth should decrease for the black hole region and increase for the de Sitter exterior. Our results will confirm this general expectation. 

\begin{figure}[h!]
    \centering
        \begin{tikzpicture}[scale=1]
            \draw[thick] (0,0) -- (0,4);
            \draw[thick] (4,0) -- (4,4);
            \draw[densely dotted] (0,0) -- (4.2,4.2);
            \draw[densely dotted] (-0.2,4.2) -- (4,0);

            \draw (4,2) node[right]{$\tau_R=0$};

            \draw[blue,thick] (4,2) -- (2.57,3.43);
            \draw[blue,thick] (1.66,2.52) -- (2.57,3.43);
            \draw[blue,dashed] (1.66,2.52) -- (0,.86);
            \draw[blue, thick] (0,1.42) -- (1.38,2.8);

            \draw[thick, color=orange, decorate, decoration={snake,amplitude=.15mm,segment length=2mm, post length=.6mm,pre length=.6mm}] (0,4) to[out=-30, in=-150] (4,4);
            \draw[thick, color=orange, decorate, decoration={snake,amplitude=.15mm,segment length=2mm, post length=.6mm,pre length=.6mm}] (0,0) to[out=30, in=150] (4,0);

            \draw[red] (4,0.18) -- (0.38,3.8);

            \draw[black, line width=0.6mm, -{Latex[length=2.5mm, width=2.5mm]}] (-0.2,4.2) -- node[above=1mm] {${U}_b$} (-.3,4.3);
            \draw[black, line width=0.6mm, -{Latex[length=2.5mm, width=2.5mm]}] (4.2,4.2) -- node[above=1mm] {${V}_b$} (4.3,4.3);

            \fill[magenta] (1.38,2.8) circle (0.05);
            \node[green] at (2.3, 3.7) {\scriptsize\textbf{ C}};
            \fill[green] (2.57,3.43) circle (0.05);
            \node[magenta] at (1.4, 3) {\scriptsize\textbf{ A}};
            \node[green] at (1.9, 2.5) {\scriptsize{ \textbf{B}}};
            \fill[green] (1.66,2.52) circle (0.05);
            \draw[dotted] (0,2) -- (4,2);
            \node[red] at (4.4, 0.3) {-$t_w$};

\draw[black, line width=0.2mm, <->, >=Latex] (0,2) -- node[left=0.6mm] {$2\Delta \tau_b$} (0,1.42);
        \end{tikzpicture}
        \label{fig:sads-a}

    \caption{Penrose diagram of the SdS$_D$ ($D > 3$) black hole interior region with a negative energy shock wave along $V_b=V_b^w \to 0^+$ at very early times. Here, we focus on geodesics that start from the static sphere at time $\tau_R = 0$. The dashed lines represent the geodesics in the absence of shock waves.
}
    \label{fig:sads-fig1}
\end{figure}
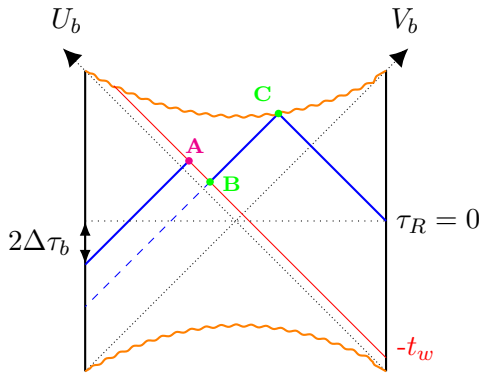

\begin{figure}[t!]
\centering

\begin{tikzpicture}[scale=1]
\draw[thick] (0,0) -- (0,4);
\draw[thick] (4,0) -- (4,4);

\draw[densely dotted] (0,0) -- (4.2,4.2);
\draw[densely dotted] (-0.2,4.2) -- (4,0);


\draw[thick, cyan] (2.3,3.4) -- (4,1.7);
\draw[thick, cyan] (1.54,2.64) -- (2.3,3.4);
\draw[thick, cyan] (1.15+0.09,2.85+0.09) -- (0,1.7);

\filldraw[blue] (4,1.7) circle (2pt);
\filldraw[blue] (0,1.7) circle (2pt);

\draw[black, line width=0.01mm, <->, >=Latex] (4,1.7) -- node[right=0.6mm] {$\Delta \tau_b$} (4,2);



\draw[black, line width=0.6mm, -{Latex[length=2.5mm, width=2.5mm]}] (4.2,4.2) -- node[above=1mm] {${V_b}$} (4.3,4.3);
\draw[black, line width=0.6mm, -{Latex[length=2.5mm, width=2.5mm]}] (-0.2,4.2) -- node[above=1mm] {${U_b}$} (-.3,4.3);

\draw[dotted] (0,2) -- (4,2);

\draw[red] (4,0.18) -- (0.38,3.8);
\node[red] at (4.4, 0.3) {-$t_w$};

\draw[dashed, black] (2,3.4) -- (0,1.41);
\draw[dashed, black] (2,3.4) -- (4,1.41);
\draw[thick, color=orange, decorate, decoration={snake,amplitude=.15mm,segment length=2mm, post length=.6mm,pre length=.6mm}] (0,4) to[out=-30, in=-150] (4,4);
\draw[thick, color=orange, decorate, decoration={snake,amplitude=.15mm,segment length=2mm, post length=.6mm,pre length=.6mm}] (0,0) to[out=30, in=150] (4,0);
\end{tikzpicture}
\label{fig:penrose-ax-b-sds-negative}

\caption{Symmetric geodesics (i.e. those with $\tau_R=-\tau_L$) in the interior black hole  
        region  of SdS deformed by a negative energy shock wave. The dashed lines represent the symmetric geodesic in the absence of a shock wave.}
\label{fig:Sdspenrose-shock wave-symmetric-b}
\end{figure}
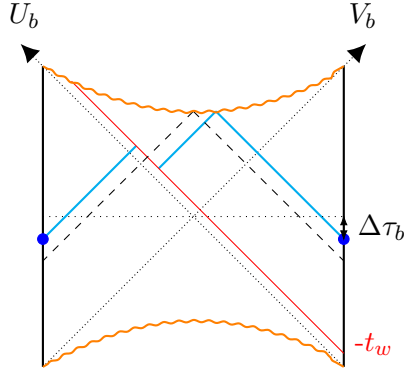

Note that the Penrose diagram of SdS spacetime in the interior black hole region looks very similar to that of the SAdS geometry, with the boundaries in this case corresponding to the static spheres instead of the AdS boundary. There is a  similar inward bending of the SdS black hole singularity with respect to the static sphere observers. The calculation of the critical times for the interior black hole region in SdS proceeds in a manner similar to our previous calculation for SAdS in Section \ref{xkutta}, except that the observer is now located at the static sphere $r = r_{\mathcal{O}} < \infty$, whereas previously, the observer was fixed at the boundary of AdS ($r = \infty$).
We will now examine the effect of a negative energy shock wave through the black hole region on the critical time.
Following similar steps as in Section \ref{xkutta}, the critical time for observers on the opposite static sphere, resulting from shock waves passing through the interior black hole region, is given by
\begin{eqnarray}\label{eq-criticaltime-U-b}
\Delta \tau_b^{\mathcal{O}} &\equiv&\gamma_{\mathcal{O}}\hs\frac{\tau_R-\tau_L}{2}=-\mathcal{T}_b^\mathcal{O} -\frac{1}{2\kappa_b ^\mathcal{O} }\log\left(1-\alpha_b\hs e^{\kappa_b(\tau_L-r^*(r_{\mathcal{O}}))}\right) .\label{SdSbhcritic}
\end{eqnarray}
The first term depends on the horizon radius and the presence of a black hole. In the absence of a black hole or in the Nariai limit, the first term vanishes. For any non-vanishing mass, it is negative.
The second term, of order $\alpha_b$, arises due to the insertion of a shock wave and is influenced by  the emission time of the null geodesic originating from the static sphere. Here, $\alpha_b$ represents the shift in the $U_b$ coordinate caused by the shock wave
\begin{align}\label{eq-alphab-Sb1}
\alpha_b&=\tilde{U}_b-U_b\simeq \frac{E}{V_b^w}  \hs \frac{dr_b}{dM}\hs C_b(r_b,r_b)=\frac{E}{V^w_b}\hs C_b(r_b,r_b)  \hs \frac{ r_b}{(D-2)\hs T_b \hs S_{b}}\,.
\end{align}
We use the usual coordinates $U_b, V_b$ for the region $V_b > V_{b}^w$ with mass $M$, and the coordinates $\tilde U_b, \tilde V_b$ for the region $\tilde V_b < \tilde V_{b}^w$ with mass $M + E$. We also point out that the shift $\alpha_b$ in the $U_b$ coordinate differs from the previous section, where we studied the shift $\alpha_c$ in the $V_c$ coordinate for shock waves in the asymptotic dS region.
Most importantly, the second term in \eqref{SdSbhcritic} contributes positively to the critical time. Negative energy shock waves can make the wormhole traversable \cite{Hirano:2019ugo} through the interior black hole region of SdS.
NEC-violating shock waves yield a positive contribution from the second term in \eqref{eq-criticaltime-U-b}, whereas NEC-satisfying shock waves always provide a negative contribution toward the critical time, as evident from \eqref{eq-criticaltime-U}.
Furthermore, we emphasize that negative energy shock waves can make the wormhole traversable \cite{Hirano:2019ugo} through the black hole region of SdS. This behavior is explicitly illustrated in figure \ref{fig:sads-fig1}, where the null geodesics originate from the right wedge at $\tau_R = 0$.

Finally, considering a symmetric pair of null geodesics, $\tau_R^\mathcal{O} = -\tau_L^\mathcal{O} = \Delta \tau_b^{\mathcal{O}}$ (see figure \ref{fig:Sdspenrose-shock wave-symmetric-b}), we substitute $\tau_L^\mathcal{O} = - \Delta \tau_b^{\mathcal{O}}$ into Equation \eqref{SdSbhcritic}. Consequently, the critical time is determined as the solution to the equation
\be 
 \Delta \tau_b^{\mathcal{O}}=-\mathcal{T}_b^\mathcal{O} -\frac{1}{2\kappa_b^\mathcal{O} }\log\left(1-\alpha_b\hs e^{-\kappa_b^{\mathcal{O}}( \Delta \tau_b^{\mathcal{O}}+\gamma_{\mathcal{O}} \hs r^*(r_{\mathcal{O}}))}\right)  .\label{sdsbhsymm-eq}
\ee
This can easily be translated to a quadratic equation for the variable $ \exp(-\kappa_b^\mathcal{O}
 \Delta \tau_b^{\mathcal{O}})$, which in turn yields only one solution with $
 \Delta \tau_b^{\mathcal{O}}$ real that corresponds to the following critical time:
\be 
 \Delta \tau_b^{\mathcal{O}}=-\mathcal{T}_b^\mathcal{O}+ \frac{1}{\kappa_b^\mathcal{O}}\log\left(\sqrt{1+\frac{\alpha_b^2}{4}}+\frac{\alpha_b}{2}\right)  . 
\label{sdsbhsymm}
\ee 
This example of symmetric geodesics explicitly demonstrates how a negative energy shock wave results in a positive contribution to the (negative) critical time, enabling the construction of a traversable wormhole through the interior black hole region of SdS. These shock waves introduce a characteristic time delay before the linear decrease of complexity ends, as well as an equal magnitude advance of the start of linear increase, shortening the complexity plateau by twice this factor. This decrease in the duration of the complexity plateau time is, as before, shorter than the corresponding scrambling time $t_w^\mathcal{O} \sim \beta_b^\mathcal{O} \log S_b$  by a $\log S_b$ factor. 

For a particular emission time $t_w$ and a given energy $E$ of the shock wave perturbations it is interesting to compare the interior black hole and exterior de Sitter time advance and delay effects. They are opposite, as they should be, producing a longer complexity plateau in the de Sitter exterior region and a shorter one in the black hole interior. Because the delays and advances are proportional to the inverse temperature, and only depend weakly on the shift $\alpha$ (through the logarithm), and because the static sphere temperature of the black hole horizon is always higher than the temperature of the cosmological horizon (except for the Nariai limit), the interior black hole shortening is expected to be smaller than the corresponding exterior de Sitter extension by a factor approximately equal to the ratio of the (static sphere) temperatures. This appears to be consistent with the fact that the total number of degrees of freedom increases as the black hole becomes smaller, as the empty, pure, de Sitter geometry corresponds to the maximum entropy state. It would be of interest to study this difference and its evolution in more detail.

\section{Summary and conclusions}

Summarizing, in this work we studied two different probes of the causal structure of the Schwarzschild-de Sitter geometry, motivated by techniques applied in the context of AdS bulk holography \cite{Fidkowski:2003nf,Festuccia:2005pi}. As a necessary condition we introduced `static sphere' Dirichlet boundary conditions, implying that we decouple the interior black hole region from the exterior de Sitter region, allowing the introduction of two bulk effective thermofield double states. It would be of interest to study the consequences of these Dirichlet boundary conditions in more detail, beyond the semi-classical quantum field theory level, especially in light of recent results on the non-uniqueness of Dirichlet boundary conditions in gravity \cite{Anninos:2024wpy, Witten:2018lgb, An:2021fcq, Anninos:2023epi, Anderson_2008}. Here we just note that small (gravitational) perturbations are not immediately expected to destabilize the system, under the assumption that the physical object, the mirror, can handle the internal pressures from the excess Hawking flux and the resulting (relative) stresses as a consequence of these perturbations. It could certainly be worthwhile to analyze this in more detail. 

With these boundary conditions in place, we first identified two sets of asymptotic quasinormal mode spectra for the interior black hole and exterior de Sitter region respectively, quoted here for the reader's convenience
\begin{align}
    \omega_n^{b,\mathcal{O}}&\approx n\pi\frac{\mathcal{T}_b^{\mathcal{O}}\pm i\frac{\beta_{b}^{\mathcal{O}}}{4}}{(\mathcal{T}_b^{\mathcal{O}})^2+(\frac{\beta_b^{\mathcal{O}}}{4})^2}\,, \label{jkchuu-conclusions}\\
       \omega_n^{c,\mathcal{O}} &\approx n\pi\frac{\mathcal{T}_c^{\mathcal{O}}\pm i\frac{\beta_c^{\mathcal{O}}}{4}}{(\mathcal{T}_c^{\mathcal{O}})^2+(\frac{\beta_c^{\mathcal{O}}}{4})^2} \, , \label{jk-c-conclusions}
\end{align}
where $\beta^{\mathcal O}_{b,c}$ are the black hole and cosmological inverse temperatures measured by the static observer, and $\mathcal{T}_{b,c}^{\mathcal{O}}$ are the critical times \eqref{mapus1} and \eqref{mapus2}, given by the bending of the black hole singularity and the cosmological future infinity, respectively. To compute them, we used 
geometric methods based on reflecting null geodesics \cite{Amado:2008hw, Law:2023ohq, Chrysostomou:2023jiv,Konoplya:2011qq}, matching 
earlier results in the pure de Sitter and Nariai limits \cite{Du:2004jt,Abdalla:2002hg}. These spectra probe the extended causal structure of the SdS geometry, and are therefore of clear holographic interest, even though they are subdominant at late times. It would be interesting to perform a more detailed study of these modes, in terms of complex geodesics \cite{Aalsma:2022eru,Chapman:2022mqd}, and how they contribute to the partition function \cite{Denef:2009kn}. 

In addition, we studied how the details of the causal structure, relative to static sphere observers \cite{Faruk:2023uzs}, affect the critical times for the interior black hole and exterior de Sitter, defined as the beginning and the end of a phase of slow evolution of the complexity parameter \cite{Anegawa:2023dad,Baiguera:2023tpt}. After having introduced and calculated these critical times for arbitrary mass of the black hole, we introduced a pair of shock waves, with vanishing total energy, and calculated the effect on the critical times. As explained, one is forced to introduce a pair of shock waves with vanishing total energy to allow for a globally consistent (patched) SdS geometry and we made the choice to emit positive energy through the cosmological horizon and negative energy through the black hole horizon, opening up both the interior black hole and exterior de Sitter wormhole region and connected to the expected thermodynamical evolution towards the maximum entropy state \cite{Aalsma:2021kle, Aalsma:2019rpt}. The critical times for symmetric geodesics in the presence of said pair of shock waves are given by
\begin{align}
 \Delta \tau_c^{\mathcal{O}}
& =\mathcal{T}_c^{\mathcal{O}}+\frac{\beta_c^{\mathcal{O}}}{2\pi}\log\left(\sqrt{1+\frac{\alpha_c^2}{4}}+\frac{\alpha_c}{2}\right),\\
 \Delta \tau_b^{\mathcal{O}}&=-\mathcal{T}_b^\mathcal{O}+ \frac{\beta_b^{\mathcal{O}}}{2\pi}\log\left(\sqrt{1+\frac{\alpha_b^2}{4}}+\frac{\alpha_b}{2}\right)  .
\end{align}
Here, $\alpha_{b,c}$ are the shifts of the Kruskal coordinates of the de Sitter and black hole regions, given by equations \eqref{eq-alphac-Sc} and \eqref{eq-alphab-Sb1}, respectively, in terms of the energy $E$ of the shock waves and the temperatures and entropies of each horizon.
In all cases we found that the effect is proportional to the inverse static sphere temperature of either the black hole interior or the de Sitter exterior, and shorter than the corresponding scrambling time. As expected, sending negative energy into the black hole shortens the complexity plateau, whereas the positive energy shock wave into the exterior de Sitter region extends it. The size of the shift is estimated to be larger in the cosmological region, which appears to be consistent with holographic considerations, but it would certainly be interesting to study this in more detail and compare to what one would expect in generic quantum chaotic systems, or more specifically SYK models that have been proposed as holographic descriptions of de Sitter space in lower dimensions \cite{Susskind:2021esx, Susskind:2021omt, VerlindeH:2023SYKdS, Verlinde:2024znh}. 

From the bulk perspective, both the quasinormal modes and the complexity parameters, next to the entropies, are interesting semi-classical probes of the SdS geometry and its extended causal structure, shedding additional light on the properties and characteristics of a holographic dual description. We hope to report on further progress in our understanding of these holographic probes of asymptotic de Sitter in the near future. 

\acknowledgments
We would like to thank Lars Aalsma, Ben Freivogel and Herman Verlinde for 
helpful discussions. JPvdS would like to thank Edward Morvan for useful discussions and feedback on a preliminary version of the draft. We would also like to thank the referee for some useful and interesting suggestions to improve the manuscript. This work is part of the Delta ITP consortium, a program of the Netherlands Organisation for Scientific Research (NWO) that is funded by the Dutch Ministry of Education, Culture and Science (OCW).

\bibliographystyle{JHEP}
\bibliography{refs}

\end{document}